\documentclass[aps, twocolumn, superscriptaddress, showpacs, nofootinbib, longbibliography]{revtex4-1}

\usepackage[utf8]{inputenc}
\usepackage[T1]{fontenc}
\usepackage{ae,aecompl} 
\usepackage{graphicx}
\usepackage{amsmath}
\usepackage{color}
\usepackage{amssymb}
\usepackage{latexsym}
\usepackage{wasysym}
\usepackage{psfrag}
\usepackage{ifthen}
\usepackage[citecolor=blue,colorlinks=true]{hyperref}
\usepackage{longtable}
\usepackage{float}
\usepackage[utf8]{inputenc}
\usepackage{lineno}

\usepackage{aas_macros}

\definecolor{bbsalmon}{rgb}{1.0, 0.47, 0.42}
\definecolor{datablue}{rgb}{0.0, 0.0, 1.0}

\begin{document}

\title{All-sky search for short gravitational-wave bursts in the third Advanced LIGO and Advanced Virgo run}

  \let\mymaketitle\maketitle
  \let\myauthor\author
  \let\myaffiliation\affiliation

\author{R.~Abbott}
\affiliation{LIGO Laboratory, California Institute of Technology, Pasadena, CA 91125, USA}
\author{T.~D.~Abbott}
\affiliation{Louisiana State University, Baton Rouge, LA 70803, USA}
\author{F.~Acernese}
\affiliation{Dipartimento di Farmacia, Universit\`a di Salerno, I-84084 Fisciano, Salerno, Italy}
\affiliation{INFN, Sezione di Napoli, Complesso Universitario di Monte S. Angelo, I-80126 Napoli, Italy}
\author{K.~Ackley}
\affiliation{OzGrav, School of Physics \& Astronomy, Monash University, Clayton 3800, Victoria, Australia}
\author{C.~Adams}
\affiliation{LIGO Livingston Observatory, Livingston, LA 70754, USA}
\author{N.~Adhikari}
\affiliation{University of Wisconsin-Milwaukee, Milwaukee, WI 53201, USA}
\author{R.~X.~Adhikari}
\affiliation{LIGO Laboratory, California Institute of Technology, Pasadena, CA 91125, USA}
\author{V.~B.~Adya}
\affiliation{OzGrav, Australian National University, Canberra, Australian Capital Territory 0200, Australia}
\author{C.~Affeldt}
\affiliation{Max Planck Institute for Gravitational Physics (Albert Einstein Institute), D-30167 Hannover, Germany}
\affiliation{Leibniz Universit\"at Hannover, D-30167 Hannover, Germany}
\author{D.~Agarwal}
\affiliation{Inter-University Centre for Astronomy and Astrophysics, Pune 411007, India}
% ...

\collaboration{The LIGO Scientific Collaboration, the Virgo Collaboration, and the KAGRA Collaboration}

\begin{abstract}
This paper presents the results of a search for generic short-duration gravitational-wave transients in data from the third observing run of Advanced LIGO and Advanced Virgo.  Transients with durations of milliseconds to a few seconds in the 24--4096 Hz frequency band are targeted by the search, with no assumptions made regarding the incoming signal direction, polarization or morphology.  Gravitational waves from compact binary coalescences that have been identified by other targeted analyses are detected, but no statistically significant evidence for other gravitational wave bursts is found. 
Sensitivities to a variety of signals are presented. These include updated upper limits on the source rate-density as a function of the characteristic frequency of the signal, which are roughly an order of magnitude better than previous upper limits. This search is sensitive to sources radiating as little as $\sim$10$^{-10} M_{\odot} c^2$ in gravitational waves at $\sim$70 Hz from a distance of 10~kpc, with 50\% detection efficiency at a false alarm rate of one per century. The sensitivity of this search to two plausible astrophysical sources is estimated: neutron star f-modes, which may be excited by pulsar glitches, as well as selected core-collapse supernova models.
\end{abstract}

\maketitle

\section{Introduction}

\label{sec:intro}

The third observing run (O3) of the Advanced LIGO \cite{LIGO-det} and Advanced Virgo \cite{Virgo-det} detectors started on April 1, 2019 and ended on March 27, 2020. During O3, tens of gravitational waves (GWs) from compact binary coalescence (CBC) were detected \cite{GWTC-2, O3IMBH, NSBH, 3-OGC}.
In addition to CBCs, there are several plausible sources of short-duration GW transients (GW bursts) that have not yet been observed,
such as core-collapse supernovae (CCSNe),  neutron star excitations, 
 non-linear memory effects, or cosmic string cusps and kinks \cite{targeted_SN_O1-2, O2magnetarbursts, S6_NS, o2_mem,  O3cosmicstring}.  
Additional source populations could exist that are yet to be predicted. For these reasons, GW burst searches capable of detecting a wide range of signal waveforms provide a unique opportunity for new discoveries.

All-sky searches look for signals arriving at any time from any sky direction.
GW searches may use signal models (targeted search) or remain agnostic about the signal morphology (generic search).
Targeted analyses include searches for CBCs \cite{GWTC-2,O3IMBH, NSBH, O3alensing} and cosmic strings \cite{O3cosmicstring}.  Generic all-sky searches look for short-duration GW transients, up to a few seconds duration \cite{allskyo2}, and for longer GW transients, up to $\sim$10$^3$ s duration \cite{allskyo2-long}. 

This paper presents results of the generic all-sky search that is sensitive to the widest range of morphologies of short duration GW bursts during O3.
The generic all-sky search is also sensitive to some CBC events \cite{allskyo2}, 
but these are not the primary targets of this analysis, and details of CBC detections during O3 are given elsewhere \cite{GWTC-2,O3IMBH}. Once the CBC candidates are excised, this search produces a null result.

This null result is interpreted in terms of sensitivities to a wide variety of generic morphologies, similarly to what was done in previous observing runs, O1 \cite{o1allsky} and O2 \cite{allskyo2}. 
The current analysis improves on previous upper limits due to improvements in detector sensitivity and a longer observation run. 
In addition, this paper includes the interpretation of results in terms of two expected astrophysical sources: CCSNe and neutron star f-modes. Since no tuning of the generic all-sky search is performed, these results should be considered conservative. 
The sensitivity of the search to five CCSNe waveform models is presented, both versus distance and for a Galactic distribution of sources. GW emission from CCSNe is expected in the frequency band below $1$~kHz. Neutron star f-modes may be excited by pulsar glitches and are expected to emit GWs in the frequency range $2$--$3$ kHz. The search sensitivity is tested for two equations of state and masses in the range $1$--$2$ $M_{\odot}$.  

The analyses described here use the final LIGO--Virgo calibration \cite{Sun_et_al_2020,O3b_LIGO_calib,O3a_Virgo_calib} and data quality \cite{O2-O3-detchar} information and their results supersede those from searches for GW bursts that were deployed in low latency during O3. The latter provided prompt public alerts for follow-up observations by other telescopes \cite{publicalertsuserguide}, analyzing near real-time data streams with preliminary calibration and data quality information.

The rest of this paper is organized as follows: Section \ref{Sec: o3run} reviews the data set used for these analyses.  Section \ref{sec:unmod} describes the two search algorithms deployed and their results (\ref{ssec:Searches}), and discusses the loudest candidate events remaining after excluding the recognized CBC candidates (\ref{ssec:marginal}).  
Section \ref{sec:interpretation} discusses the sensitivity of this all-sky search and sets new rate-density limits for transient GW events other than CBC, as well as the sensitivity to CCSN models and to neutron star f-modes.  Finally,  Section \ref{Discussions} summarizes the results and implications from this minimally-modeled search for GW transients.

\section{O3: The Third Advanced-Detector Observing Run}\label{Sec: o3run}

\subsection{Data set}
\label{ssec:data}

\begin{figure}[t!]
  \centering
    \scalebox{0.9}{\includegraphics{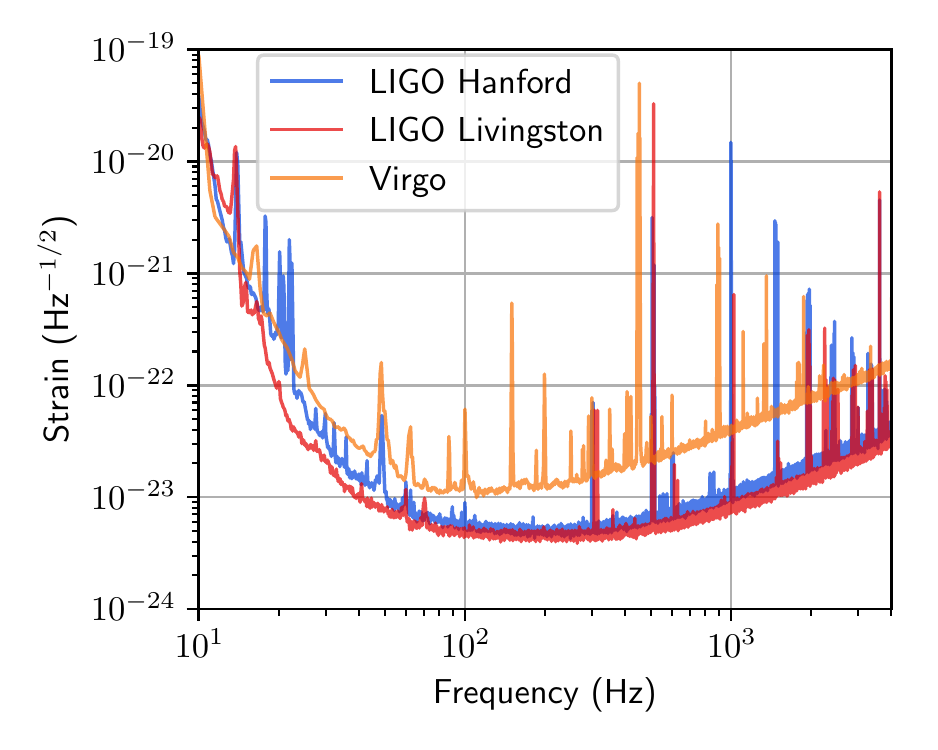}}
  \caption{Representative amplitude spectral density of the three detectors’ strain sensitivity (LIGO Livingston 5 September 2019 20:53 UTC, LIGO Hanford 29 April 2019 11:47 UTC, Virgo 10 April 2019 00:34 UTC).}
  \label{Fig.O3a_ASD}
\end{figure}

The O3 data set extends from April 1, 2019 to March 27, 2020. A commissioning break between October 1, 2019 and November 1, 2019 separates the first 6-month epoch (O3a) of the observing run from the second epoch (O3b). 
Figure \ref{Fig.O3a_ASD} shows typical spectral sensitivities of the detectors.
The Hanford--Livingston (HL) network is analyzed during times where these two detectors operated in coincidence. In addition, results for the Hanford--Virgo network (HV) and the Livingston--Virgo network (LV) are presented for times when data from either of the LIGO detectors is not available.  See Section \ref{sec:unmod} for an explanation of
why the two detector network is preferred over the three detector Hanford--Livingston--Virgo (HLV) network for this search.

During the six months of O3a, 130.2~days of data were collected at Hanford, 138.5~days of data were collected at Livingston, and 139.5~days of data were collected at Virgo. The amount of data actually analyzed is reduced by requiring
coincidence between two detectors, removing poor periods of data quality as described in Section \ref{ssec:DQ}, and requiring at least 200 seconds of continuous observation-quality data.  This results in the following total amounts of analyzed
 data: 104.9~days for the HL network, 14.8~days for the HV, and 25.6~days for the LV network.

During O3b, data were collected for 115.7~days at Hanford, for 115.5~days at Livingston, and for 113.2~days at Virgo. The actual analyzed data amounts are 93.4~days for HL, 17.8~days for HV, and 14.8~days for the LV network.

The calibration uncertainties for the LIGO detectors in the 20--2000 Hz frequency range are <7\% in amplitude, <4$^{\circ}$ in phase, <1~$\mathrm{\mu}$s in timing for O3a \cite{Sun_et_al_2020}, and <12\% in amplitude, <10$^{\circ}$ in phase, <1~$\mathrm{\mu}$s in timing for O3b \cite{O3b_LIGO_calib}. The calibration uncertainties for Virgo in most of the 20--2000 Hz frequency range are <5\% in amplitude, <2$^{\circ}$ in phase, and <10~$\mathrm{\mu}$s in timing for all of O3 \cite{O3a_Virgo_calib,Rolland:2020vna}. These uncertainties are not expected to have a significant impact on the search presented here. However, they can contribute to the systematic uncertainties associated with the efficiency numbers quoted in Section \ref{sec:interpretation}. The O3a GW strain data used in this paper is part of the O3a Data Release through the Gravitational Wave Open Science Center~\cite{GWOSC}, and can be found at~\cite{GWOSC_O3a}. 

\subsection{Data quality}
\label{ssec:DQ}

The LIGO and Virgo detectors are affected by various sources of terrestrial 
noise that can interfere with the detection of GWs \citep{O3performance,Abbott_2016}.  In addition to the primary channel recording GWs, the
interferometers use a large number of auxiliary channels that observe either the external environment \citep{envnoise,PEM}, or the interferometer itself.
 Through the use of auxiliary channels, it is possible to substantially reduce the impact of noise transients by 
discarding (vetoing) a small percentage of observing time during which noise contamination is present \cite{dqmitigation}.  A brief discussion of some of the most relevant data quality issues is presented in this section, but much more detail on these 
issues and their mitigation can be found in \citep{O2-O3-detchar}.

To address specific data quality problems, tens of different data quality vetoes defining times to be removed from the search are constructed and applied to the analyses described in
 this paper.   The effectiveness of each data quality veto is determined based on the ratio of the percentage of glitches removed to amount of observation time vetoed. The most significant data quality issues successfully discarded by these vetoes are high signal-to-noise ratio (SNR)
glitches associated with light intensity dips in both LIGO interferometers, radio 
frequency beatnotes (also known as Whistles), and a single half-hour period of
extremely rung-up violin mode resonances of the LIGO Hanford suspension system.
An additional stage of automated statistical vetoes using the
 hveto \cite{Smith_2011} algorithm is subsequently applied using the same
procedure as in O2 \cite{allskyo2}.  Hveto uses a hierarchical method to produce a ranked list of statistically significant vetoes
generated by applying a specific list of SNR thresholds and time windows to a
subset of LIGO's auxiliary channels.
 Between 1\% and 2\% of the total observation time per interferometer is discarded due to data quality issues, with precise breakdowns provided in \cite{O2-O3-detchar}.  A complete list of vetoes used in this search with brief
descriptions of each is given in \cite{o3dataflags}.

Unfortunately, these vetoes do not suppress all non-astrophysical features of the data.  As interferometer 
sensitivity has improved, light scattering has become more prominent at low 
frequencies \citep{Soni_2021,Soni_2021b}.  Scattering noise was significantly reduced
 in the latter part of the run, but it remained a prominent feature throughout much of O3, especially during periods of high anthropogenic or seismic activity.  Because of the large amount of time with light scattering present and the lack of straightforward and consistent auxiliary channel witnesses, most light-scattering glitches are not vetoed.

Another prominent noise type that is not vetoed by standard methods are Blip glitches \cite{Cabero_2019}.  These have 
recurred in both LIGO interferometers throughout the advanced detector era. 
 Blips are 
short-duration, low quality factor ($Q$) glitches which occur at the rate of several 
per day.  As these Blips do not have clear auxiliary witnesses or known origin, 
and are not clearly morphologically distinct from some astrophysical models of 
interest, they must be handled by the search algorithms themselves.
During O3 a new population of loud single-pulse Blip-like 
glitches was observed.  The origin of these glitches is not known.  
See Section \ref{ssec:cWB} for more details on the handling of this glitch class.

\section{Unmodeled GW transient analyses} 
\label{sec:unmod}

Using the three-detector HLV network generally enables more accurate reconstruction of both the structure of the GW signal and its sky location than is possible with a two-detector network. However, for purposes of detection, the sensitivity of the HLV network is not better than the HL network for the O3 analyses described in this paper.  The generic all-sky search for GW bursts cannot rely on assumptions about the GW polarization state. Since the two LIGO interferometers are nearly co-aligned and therefore sensitive to similar linear combinations of the GW polarization components over most sky directions, Hanford and Livingston 
generally detect a given GW with comparable SNRs.
Virgo, by contrast, typically senses a different linear combination of GW polarizations.  In O3 the LIGO interferometers have better sensitivity than Virgo
 (see Figure~\ref{Fig.O3a_ASD}), and for many source directions the difference in detector orientation enhances this disparity. 

In addition, there is a negligible loss in detection efficiency when
narrowing the analysis of HL-only data to search for the GW polarization
projection that best matches the network from each sky direction.
This allows implementing stricter requirements on the signal coherence
between the Hanford and Livingston detectors and results in a more
effective rejection of noise transients.
This advantage is not shared by analyses of networks involving Virgo due to its misalignment with the LIGO detectors. To make full use of Virgo data, the analysis has to either open the search to both GW polarization components over the sky, or relax the requirements on the signal coherence between participating detectors.  The distribution of non-Gaussian noise outliers present in all detectors in O3 is thus more effectively mitigated in coherent analyses of the HL network than in analyses with networks including Virgo, and this affects the resulting detection efficiency. The analyses described
in this paper therefore use the HL network rather than HLV because we are interested in maximizing detection probability.

The search for short GW bursts is sensitive to CBC sources, especially binary black hole coalescences \cite{allskyo2}, and hence a fraction of them are found by the analyses presented here. The discussion of the astrophysical properties and implications of the detected CBC events is the subject of other LIGO Scientific and Virgo Collaboration catalog papers (see \cite{GWTC-2} for O3a results). Search results in this paper
initially include GWs from CBCs, but known CBC events are excised in a subsequent
step, and discussion here is limited to candidate events that are not found by targeted searches for such sources.

\subsection{Search algorithms}\label{ssec:Searches}

In order to make the results of the search more robust, two independently developed search algorithms are deployed: coherent WaveBurst (cWB) and BayesWave (BW). The cWB algorithm is used to analyze the entire dataset. BW is computationally more intensive, thus it is only used to follow up a subset of the dataset identified by cWB in 
order to provide a partly independent measurement of the candidates' significance. Both of these algorithms and their results are described below. 

\subsubsection{Coherent WaveBurst} \label{ssec:cWB}

Coherent WaveBurst is an algorithm based on the maximum-likelihood-ratio statistic over all sky directions applied to excesses of signal power in the time--frequency domain representation of the strain data from the network of detectors \cite{cwb2g,SOFTX,cwb_repo}. The analysis uses the Wilson–Daubechies–Meyer wavelet transform at various time--frequency resolutions \cite{WDM}. Multiple resolutions allow adaptation of time--frequency characterization to the signal features.  Coherent WaveBurst is routinely used in LIGO--Virgo searches and reconstruction of GW transients \cite{allskyo2,o1allsky}.   

In this work the low and high frequency parts of the parameter space are separately covered by two analyses.  The same procedure was also done for O1 \cite{o1allsky} and O2 \cite{allskyo2}.  
The clusters of wavelets which fall above the noise floor of the detectors and pass the internal thresholds of the pipeline are referred as \textit{triggers}.  

The low-frequency analysis covers the frequency range between $16$--$1024$~Hz.  
Triggers with mean reconstructed frequency
below  $24$~Hz and $32$~Hz are rejected for O3a and O3b, respectively,  to avoid contamination from loud and frequent low-frequency glitches.  
The HV and LV networks are analyzed only when one of the LIGO interferometers is unavailable, i.e. there is no overlap in time of data set for any of the networks considered and the livetimes for each network are mutually exclusive.

The requirement on the signal coherence across detectors is implemented as a threshold on the network correlation coefficient (referred to as $c_c$ in \cite{cwb2g}), which is the fraction of coherent energy in the network of detectors. 
After inspection of the overall performance over the set of signal models listed in Table~\ref{Tab.O3hrss50}, triggers are required to pass $c_c$ thresholds of 0.8 for the analysis of the HL network and 0.5 for the HV and LV networks, since Virgo is not co-aligned with the two LIGO detectors.

The triggers obtained after passing the frequency rejection and network correlation coefficient threshold are further divided into three different, mutually exclusive bins, referred to as LF1, LF2 and LF3.  The choice of the bins is based on the background triggers' morphologies,  and the goal is to isolate background triggers that are loud and frequent to a small part of the parameter space.  LF1 contains triggers with most of the signal energy confined to a single oscillation. In O3 a population of such short-duration Blip glitches dominates the tail of the background trigger distribution and hence 
a new bin is introduced in the O3 search to confine these glitches (see Section \ref{ssec:DQ}).  
LF2 contains the remaining triggers that are characterized by $Q \leq 3$, also resembling Blip glitches. LF3 contains the higher $Q$ low-frequency triggers and shows the cleanest background distribution. Unlike O1 and O2, non-stationary spectral lines do not contribute significantly to the background in O3.

The low-frequency cWB analysis is performed separately for O3a and O3b.  
The background distribution of triggers is calculated by time-shifting the data of one detector with respect to the other detector by an amount that breaks any correlation between detectors for a real signal.  The HL network is time-shifted to obtain total background livetime of around 2000 years.  For the HV and LV networks, around 1000 years of background are generated using all coincident data.  The use of full coincident time for the HV and LV networks is necessary because the exclusive livetime is not sufficient to produce such large background statistics.  

The high-frequency analysis covers the frequency range 1024--4096~Hz.  The analysis is carried out in the frequency band 512--4096~Hz  but only triggers with mean reconstructed frequencies are $ \geq 1$~kHz are kept.   For this analysis only the HL network is considered, as Virgo is significantly less sensitive than the LIGO interferometers in the high-frequency region (a factor of $\sim$5 above 1000~Hz, see Figure~\ref{Fig.O3a_ASD}). Similarly to the low-frequency analysis, a network correlation coefficient threshold of $0.8$ is used for the high-frequency part of the analysis.  No division of background triggers into analysis bins is required for this analysis.  However,  during the first part of O3 run until May 16, 2019
there were background triggers dominating the tail with central frequency $f_0 > 3400$~Hz;  for this part of the run only the triggers with central frequency $ \leq 3400$~Hz are admitted.  The full frequency range is considered for all times from 
May 16 onward. As a result, the high-frequency cWB analysis is divided into three parts, the first two parts are in O3a (before and after May 16, 2019, see above),  and the third part corresponds to all of O3b.  Total background livetimes of around 1000 years are produced for O3.  

The significance of each trigger is calculated by comparing the coherent network SNR $\eta_c$ \cite{cwb2g} with the background distribution of the bin and the network to which the trigger belongs.  The \textit{inverse false alarm rate} (iFAR) is calculated for each observed trigger.  The iFAR for the low-frequency analysis is penalized by a trials factor of 3 corresponding to the three analysis bins LF1, LF2 and LF3.
The criteria for a significant detection of an event is set at iFAR $\geq 100$ years.

\begin{figure}[t!]
  \centering
  \scalebox{0.7}{\includegraphics{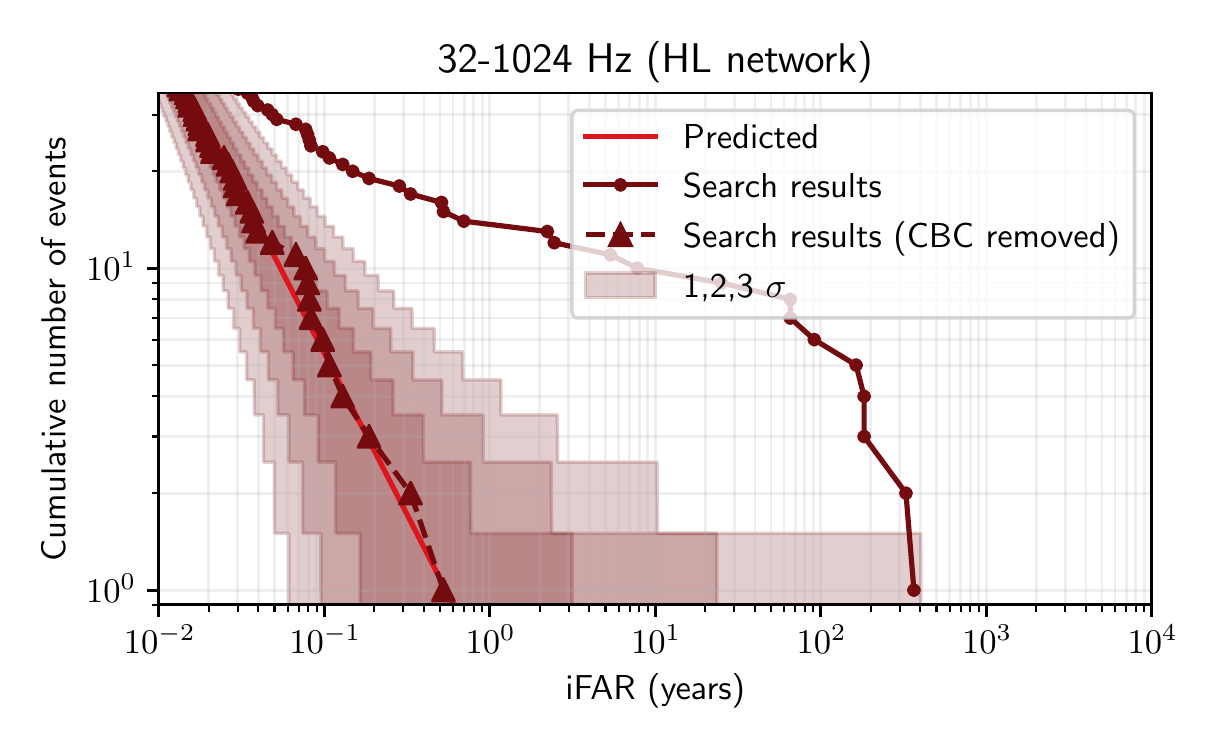}}
    \scalebox{0.7}{\includegraphics{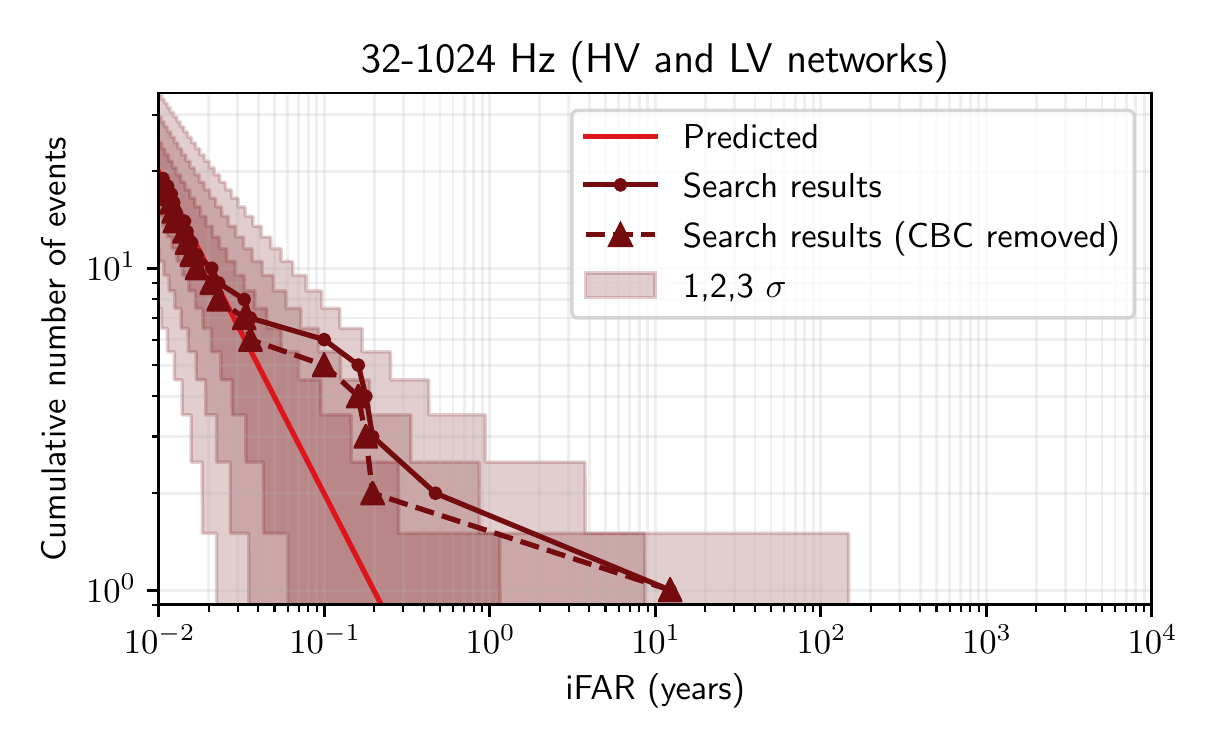}}
  \caption{Cumulative number of events versus inverse false alarm rate (iFAR) found by the cWB low-frequency analysis using all O3 data for the HL network (top panel), and the HV and LV networks combined (bottom panel).  Circular points show results for all data and triangular points show after times around all known compact binary coalescence sources have been excised.  The solid line shows the expected mean value of the background, given the analyzed time. The shaded regions show the 1, 2, and 3 $\sigma$ Poisson uncertainty regions.}
  \label{Fig.O3_cWB_FAR_LF}
\end{figure}

The analysis results for the cWB low-frequency region are shown in Figure \ref{Fig.O3_cWB_FAR_LF}.   
The loudest candidate event in the HL network after excluding known CBCs \cite{GWTC-2} occurred at UTC 2019-09-28 02:11:45.
This candidate has an iFAR of  $0.53$~years. The second most significant candidate in this network occurred at UTC time 2019-08-04 08:35:43, with an iFAR of $0.19$~years.  
The loudest candidate for the HV and LV networks is an HV event at UTC time 
2019-04-30 00:49:32, with an iFAR of $12$~years.  Though none of these meet the iFAR threshold of 100 years for a detection,  investigations into these loudest remaining candidates are discussed further in Section \ref{ssec:marginal}.  

\begin{figure}[t!]
  \centering
    \scalebox{0.7}{\includegraphics{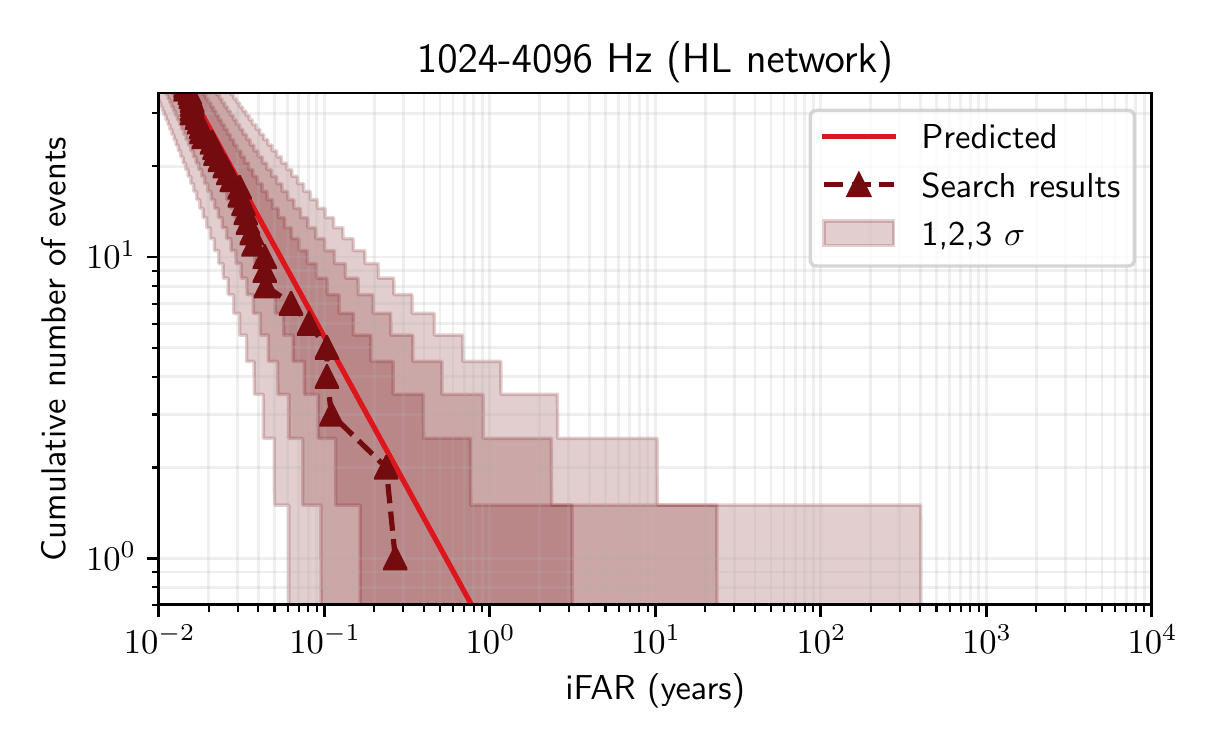}}
  \caption{Cumulative number of events versus inverse false alarm rate (iFAR) found by the cWB high-frequency analysis (triangular points) using all O3 data for the HL network (Virgo is not used for high-frequency search). The solid line shows the expected mean value of the background, given the analyzed time. The shaded regions show the 1, 2, and 3 $\sigma$ Poisson uncertainty regions.}
  \label{Fig.O3_cWB_FAR_HF}
\end{figure}

The results for the high-frequency cWB analysis are shown in Figure \ref{Fig.O3_cWB_FAR_HF},  the loudest event has an iFAR of $0.3$~years.

\subsubsection{BayesWave}
\label{ssec:bayeswave}

BW \cite{Cornish:2014kda, bw_repo, BayesWaveMethodsIII} is a Bayesian algorithm modeling both GW signals and non-Gaussian noise transients as sums of sine-Gaussian wavelets. The number of wavelets used is marginalized over using a trans-dimensional Reversible Jump Markov Chain Monte Carlo algorithm. The detection statistic used is the natural logarithm of the signal-to-glitch Bayes factor ($\ln\mathcal{B}_{\mathcal{S},\mathcal{G}}$), i.e.~the Bayes factor between the signal model consisting of Gaussian noise and an astrophysical signal coherent across detectors; and the glitch model, which describes the data as Gaussian noise and glitches modeled independently in each detector. Thus a positive $\ln\mathcal{B}_{\mathcal{S},\mathcal{G}}$ indicates that the presence of a GW signal is favored, while a negative $\ln\mathcal{B}_{\mathcal{S},\mathcal{G}}$ shows support for the event being a glitch.

Due to the trans-dimensional sampling it requires, analyzing the entire O3 dataset with BW is computationally prohibitive. Thus BW is used as a follow-up to the cWB pipeline, similarly to previous observing runs \cite{o1allsky, allskyo2}. By doing so an additional
measurement of iFAR for the candidates followed up by BW is acquired, thus making the search presented in this paper more robust against potential shortcomings of individual algorithms.
BW followed up cWB candidates in the low-frequency analysis, treating all the search bins as a single bin, and using a threshold of $\eta_c = 9.90$. BW uses the same background data set as cWB from time slides.

\begin{figure}[t!]
  \centering
  \scalebox{0.7}{\includegraphics{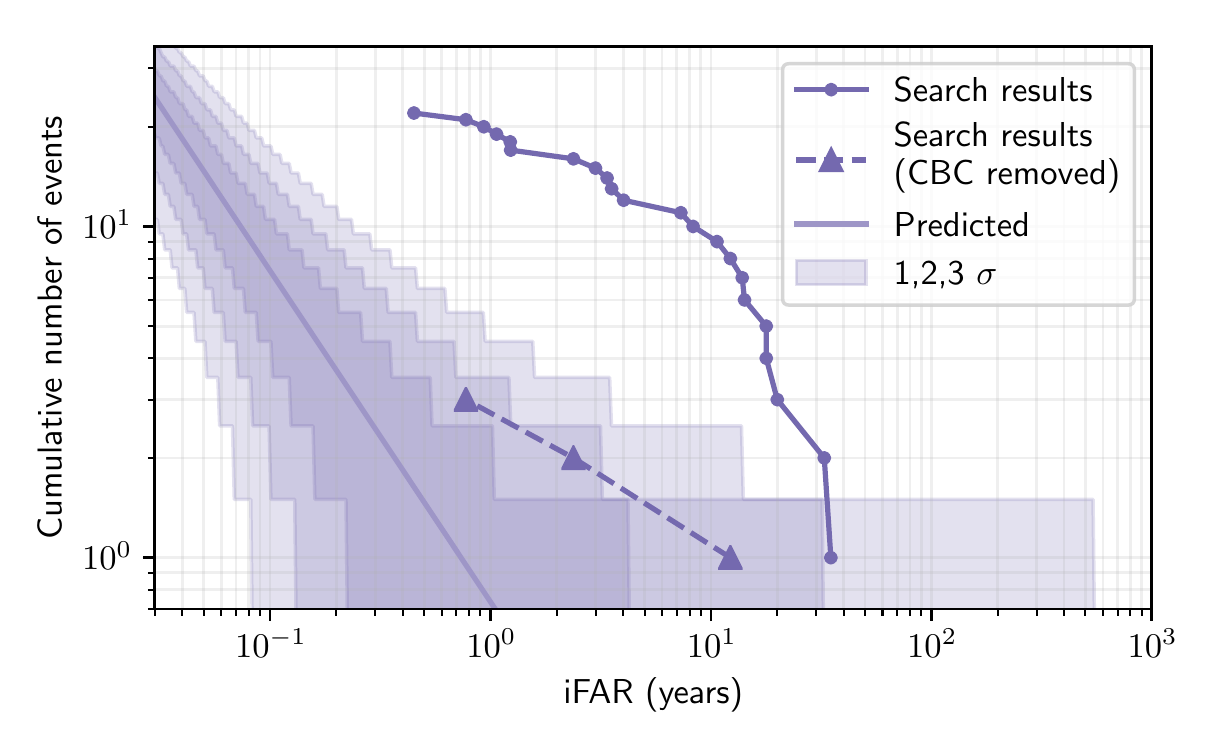}}
  \caption{Cumulative number of events versus inverse false alarm rate (iFAR) found by the BW follow-up to the cWB low-frequency analyses using all O3 data (circular points) and after times around all compact binary coalescence sources have been excised (triangular points). The solid line shows the expected background, given the analyzed time.
  The shaded regions show the 1, 2, and 3 $\sigma$ Poisson uncertainty regions.}
  \label{Fig.O3_BW_FAR}
\end{figure}

A total of 22 cWB candidates are above the $\eta_c$ threshold, 19 of which are known CBC candidate events described in recent or future publications. This is fewer than found by cWB, because not all CBC candidates passed the BW follow-up threshold.
The combined results from all detector networks are shown in Figure \ref{Fig.O3_BW_FAR} in terms of the cumulative distribution of their iFAR values. 
The three candidate events remaining after removing the known CBC candidate events
are discussed in the previous section.
None of these is found with high enough significance in BW to be considered a likely GW event.  Section \ref{ssec:marginal} discusses these candidate burst events.

\subsection{Candidate events}
\label{ssec:marginal}

\subsubsection{Surviving non-CBC candidates}

The three non-CBC candidate events with $\eta_c$ values above 9.90, a high enough value to trigger BW follow-up, are discussed individually below. They are identified by the time at which they occurred. In each case, the statistical significance
is not high enough to claim the candidate as a GW event.  Though none of these candidates are vetoed by data quality procedures, data quality concerns for each case are discussed.

{\bf 2019-09-28 02:11:45 UTC}  The most significant HL cWB candidate has an inverse false alarm rate of 0.53~years in cWB all-sky and 0.8~years in the BW follow-up.  This initially appeared in the low-latency CBC-focused cWB 
analysis but was manually rejected in near-real time as most probably being caused by a glitch in the Livingston detector \cite{GCNS190928c}.  It does not pass  signal consistency cuts specific to the version of that 
search focused on CBCs, described in \citep{GWTC-2}, 
but remains in the more general burst 
analysis at lower significance.  Instrumental investigations into possible
origins focused on magnetic noise in the station at the end of Hanford's X-arm, but magnetic coupling was ruled out as a significant contributor to the power of the putative signal. 
The morphology in the Livingston detector resembles 
Tomte glitches \citep{O3performance,gravityspy} appearing at other times,
while there is little power in the Hanford detector. 
The significant difference in the relative amplitude between Hanford and 
Livingston would mean that, if astrophysical, this candidate event would have to originate from the $\sim$5\% of the sky where Hanford has negligible sensitivity but Livingston's sensitivity is significant.

{\bf 2019-08-04 08:35:43 UTC}  The second most significant low-frequency HL cWB candidate, 
at an iFAR of 0.19 years, was also initially identified in a low-latency cWB targeted search for binary black hole coalescences.  BW follow-up finds an iFAR of 12.2~years, making this the most significant non-CBC outlier in that analysis.
It occurred less than a second after an SNR $\sim$60 series of glitches in Livingston, which are themselves too loud to be astrophysical.  These glitches morphologically resemble the Repeating Blips class of glitches \citep{gravityspy} occurring at other times in both LIGO interferometers.  Its close proximity to these glitches makes it impossible to
discount an instrumental origin, though it is not vetoed by
any auxiliary witness channel.  As a follow-up study to the low-latency search, BW was used to model the glitches occurring just before the candidate event, and that model was subtracted from the data in order to produce a de-glitched data stream \cite{BayesWaveMethodsIII}.  It was found 
that this glitch subtraction lowered the SNR but had negligible effect on the reconstructed morphology of the candidate.

{\bf 2019-04-30 00:49:32 UTC}  An additional candidate is identified in the HV O3a cWB search, a less sensitive network than HL, at an iFAR of 12.29~years. The BW follow-up gives an iFAR of 2.4~years for this trigger.
The presence of Blip glitches in Hanford less than a second prior to the candidate and the resemblance to a Blip glitch in the Hanford interferometer lead to similar data quality concerns as the previous trigger. 

\subsubsection{Low-latency-only candidates}

In the low-latency search described in Section \ref{sec:intro}, public alerts were
generated for burst search candidates with significance exceeding an iFAR of 4 years. Two candidate events crossed this significance threshold in the low-latency cWB search, but do not appear in the version of the analysis presented in this
paper, as explained below.

{\bf S191110af} This was a high-frequency ($\sim$1780~Hz) HL cWB candidate that 
generated a public alert \citep{GCNS191110af} based on its significance in the low-latency cWB analysis.
Follow-up of the candidate shortly after it was identified indicated that it
was due to a faulty piezoelectric transducer at Hanford.
This candidate event does not appear in the analysis described in this paper, as times 
strongly affected by this noise were vetoed \citep{O2-O3-detchar}.  It is no longer of astrophysical interest.

{\bf S200114f} This HL candidate generated a public alert \citep{GCNS200114f} based on its significance in the low-latency unmodeled cWB all-sky search, but is not found in the analysis as described in this paper because it fails an internal cWB consistency cut (the network correlation coefficient $c_c < 0.8$, see Section \ref{ssec:cWB}) requiring the
signal to be correlated between the two LIGO detectors. 
It is further discussed in the O3 intermediate mass black hole search paper \cite{O3IMBH}. 

\section{Astrophysical interpretation of the results}
\label{sec:interpretation}

In order to place the search results in an astrophysical context, it is necessary to measure detection efficiency for plausible signals. This is accomplished by injecting simulated signals (via software) into the detector data and recovering them using the search methods described in previous sections. The pipelines’ abilities to recover a broad range of transient signals can be tested by this method.   
These transient signals include a set of {\it ad hoc} waveforms as well as astrophysically motivated waveforms from CCSNe and neutron star f-modes.
The sensitivity of the search to these simulated signals is described in this section.  Only the HL network is used for quoting sensitivities, as the other network pairs provide sensitivities which are at least a factor of 2 worse in amplitude.  The sensitivities quoted in this section follow the criterion for significant detection of iFAR $\geq$ 100 years.  

\subsection{Sensitivity to generic signal morphologies}
\label{ssec:generic_sensitivity}

As the pipelines are able to detect GWs from a range of potential astrophysical sources, a set of {\it ad hoc} waveforms comprising a wide range of different morphologies are used to estimate the search sensitivity to generic signals.  The waveform families used here are sine-Gaussian wavelets (SG), Gaussian pulses (GA), and band-limited white-noise bursts (WNB).  The SG signals are defined by the central frequency $f_0$ and quality factor $Q$, which determine the signal's duration. The GA signals are described by the duration of one standard deviation $\tau_{\mathrm{GA}}$.  The WNB signals are described by their lower frequency bound $f_\mathrm{low}$, bandwidth $\Delta f$, and duration $\tau_{\mathrm{WNB}}$.  Further details on these waveform morphologies can be found in the S6 short duration all-sky search  \cite{s6Burst}.  
These {\it ad hoc} signals are injected in the network of detectors over a range of amplitudes, which are expressed in terms of the root-mean-squared strain amplitude ($h_\mathrm{rss}$) given by 
\begin{equation}
h_{\mathrm{rss}} = \sqrt{\int_{-\infty}^{\infty} \left(h^2_+(t) + h^2_{\times}(t) \right)  \mathrm{d}t},
\end{equation}
where $h_{+}$ and $h_{\times}$ are the components of the signal polarizations in the source frame.  \\

There are differences in the distribution of extrinsic parameters for the SG and GA injections with respect to the O2 search~\cite{allskyo2}.  For the SG and GA waveforms in O3 the simulated signals are injected 
over a grid of maximum strain values given by $h_\mathrm{rss} = \left(\sqrt{3}\right)^N 5\times10^{-23}$ $\text{Hz}^{-1/2}$, where $N$ ranges from 0 to 8. 
The strain distribution for the O2 search was uniform in the square of the signal distance.  Similarly to O2, the simulated signal sources are drawn from a uniform distribution in solid angle over the sky.  The polarization for GA waveforms is linear,  whereas the SG waveforms use both elliptical SG, which are uniform in cosine of the inclination angle of the source,  and circular SG, which assume an optimally oriented source.  The inclination angle is defined by the angle between the total angular momentum vector and the line of sight.  In order to have a direct comparison of sensitivity between the observing runs, the same set of WNB waveforms as described in~\cite{allskyo2} are injected into O3 data. 

The $h_\mathrm{rss}$ values at which 50\% of signals are detected with an ${\rm iFAR}\geq 100\ {\rm years}$ for each waveform are given in Table~\ref{Tab.O3hrss50}. Calibration uncertainities affect the results to at most 10\% as discussed in Section \ref{ssec:data}. Results for the SG waveforms are given only for the circular SG, which is the best case scenario.  The results show an imbalance in the sensitivity of the cWB and BW pipelines for SG waveforms. This is due to the fact that the detection statistic $\ln\mathcal{B}_{\mathcal{S},\mathcal{G}}$ used by BW scales linearly with the number of wavelets used in the reconstruction~\cite{Kanner:2015xua,Lee:2021}.  Because SG and GA waveforms can be accurately reconstructed using a single wavelet, BW is less sensitive to these particular signals.  For O3 the sensitivity to GA is worse compared to O2,  this is mainly due to the population of Blip glitches during O3 that resembled GA injections, and are isolated in a dedicated bin as described in Section \ref{ssec:cWB}.  

\begin{table}[!htb]
\centering
\scriptsize
\begin{tabular}{ c c c  c c}
\hline
\hline
   & \multicolumn{2}{c }{O3a} & \multicolumn{2}{c}{O3b} \\
\hline
\textbf{Morphology} & \textbf{cWB}  & \textbf{BW}& \textbf{cWB}  & \textbf{BW} \\
\hline
\textbf{Gaussian pulses (linear)}\\

$\tau_{\mathrm{GA}} = 0.1$ ms & 18.1 & - & 8.2 & - \\
%\hline
$\tau_{\mathrm{GA}} = 2.5$ ms & 25.2 & -  & 10.5 & -  \\
%\hline
\textbf{Sine-Gaussian wavelets (circular) }\\
%\hline
$f_0 = 70$ Hz, $Q = 3$ & 1.1 & $>40$ & 1.1 & $>40$  \\
%\hline
$f_0 = 70$ Hz, $Q = 100$ & 1.0 & $>40$  & 1.0 & $>40$ \\
%\hline
%$f_0 = 153$ Hz, $Q = 8.9$ & 0.837 & $>40$ & 0.852 & $>40$  \\
%\hline
$f_0 = 235$ Hz, $Q = 100$ & 0.8 & 2.5 & 0.8 & 3.7 \\
%\hline
$f_0 = 554$ Hz, $Q = 8.9$ & 1.0 & $>40$ & 1.1 & $>40$  \\
%\hline
$f_0 = 849$ Hz, $Q = 3$ & 1.5 & $>40$ & 1.6 & $>40$  \\
%\hline
$f_0 = 1304$ Hz, $Q = 9$& 1.9 & - & 1.9 & -  \\
%\hline
$f_0 = 1615$ Hz, $Q = 100$ & 2.2 & -  & 2.4 & -  \\
%\hline
$f_0 = 2000$ Hz, $Q = 3$ & 3.2 & - & 3.1 & -  \\
%\hline
$f_0 = 2477$ Hz, $Q = 8.9$ & 3.8 & - & 3.7 & -  \\
%\hline
$f_0 = 3067$ Hz, $Q = 3$ & 5.6 & - & 5.0 & -  \\
%\hline
\textbf{White-noise bursts}\\
%\hline
$f_{\mathrm{low}} = 100$ Hz, $\Delta f = 100$ Hz, $\tau_{\mathrm{WNB}} = 0.1$ s & 0.9& 2.6 & 1.0 & 3.4  \\
%\hline
$f_{\mathrm{low}} = 250$ Hz, $\Delta f = 100$ Hz, $\tau_{\mathrm{WNB}} = 0.1$ s & 0.9 & 2.2 & 1.0 & 3.5  \\
%\hline
$f_0 = 750$ Hz, $\Delta f = 100$ Hz, $\tau_{\mathrm{WNB}} = 0.1$ s & 1.5 & 2.8 & 1.5 & 3.9  \\
\hline
\hline
\end{tabular}
\caption{The $h_\text{rss}$ values (in units of $10^{-22}$ $\text{Hz}^{-1/2}$) for which 50\% detection efficiency is achieved with an iFAR of 100 years for each of the injected signal morphologies.  The SG waveforms reported in this table have circular polarization. ``$>40$'' indicates that 50\% detection efficiency is not achieved for the maximum $h_\mathrm{hrss}$ used in this injection set, and ``-'' denotes waveforms not analysed by BW.}\label{Tab.O3hrss50}
\end{table}

\begin{figure}[t!]
  \centering
 \scalebox{0.95}{\includegraphics[width=0.5\textwidth]{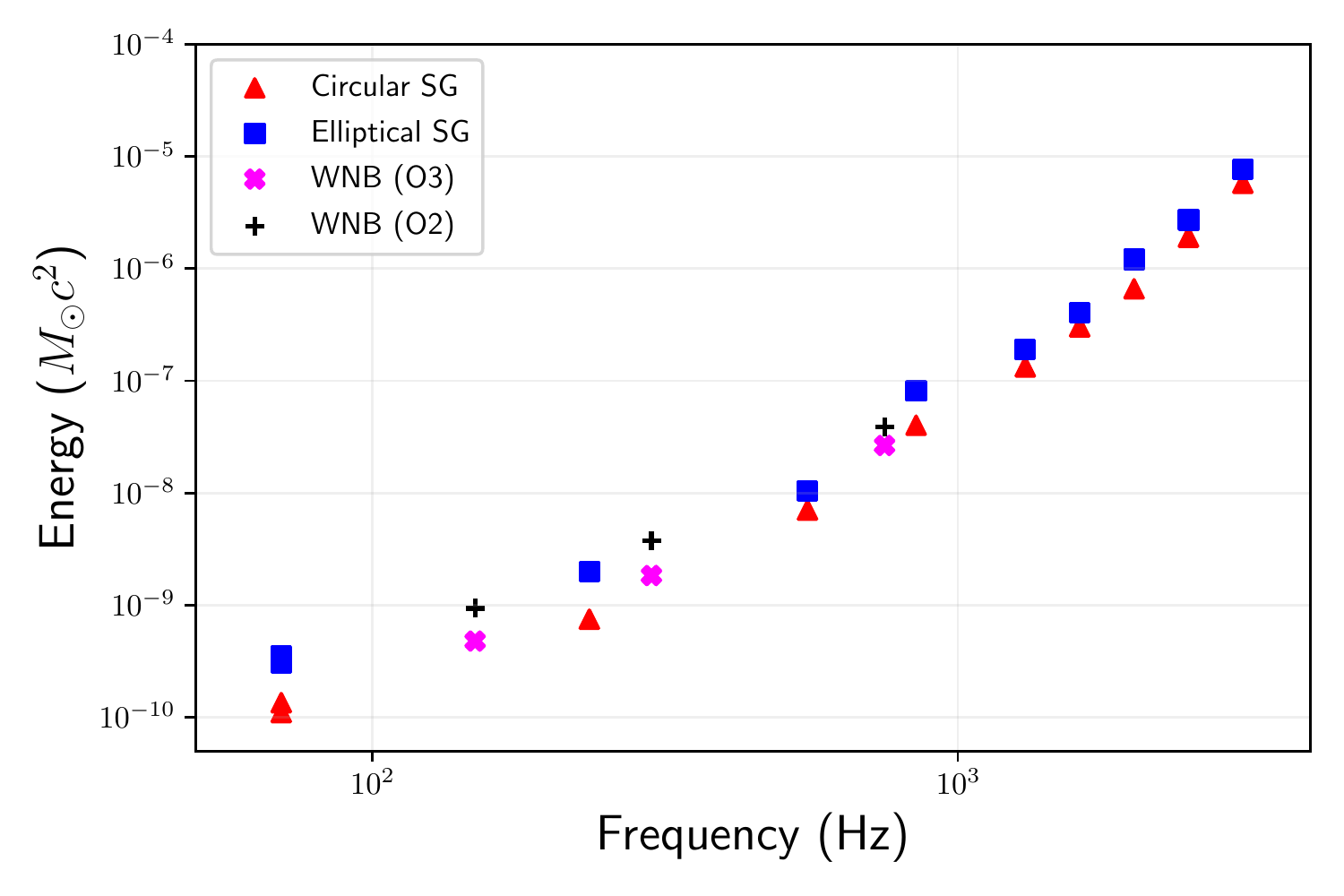}}
  \caption{
The GW emitted energy in units of solar masses that correspond to a 50\% detection efficiency at an iFAR of $\geq$ 100 years,  for a source emitting at 10 kpc. The waveforms represented here include all of the circular SG and WNB injections as given in Table~\ref{Tab.O3hrss50} using $E_{\mathrm{GW}}=E_{\mathrm{GW}}^{\mathrm{iso}}$.  The SG waveforms with uniform distribution in cosine of inclination angle (elliptical SG) are also reported  using $E_{\mathrm{GW}}=E_{\mathrm{GW}}^{\mathrm{rot}}$.  Only cWB results are presented for O3 as it is the most sensitive pipeline for the injection set used here.  The same results for O2 are also shown for comparison for the WNB waveforms.}  
    \label{Fig.O3EvF}
\end{figure}

The detection efficiencies obtained can be used to determine the typical amount of energy emitted in GWs needed for a detection.  This is done assuming a standard-candle source at a distance of $r_0=10$ kpc radiating GWs isotropically at a central frequency of $f_0$.  The amount of energy radiated assuming the signal to be narrow band is then given by \cite{s6Burst}
\begin{equation}
  \label{Eq:Egw}
  E_{\mathrm{GW}}^{\mathrm{iso}} = \frac{\pi^2 c^3}{G}r_0^2 f_0^2 h_{\mathrm{rss}}^2.
\end{equation}
This equation is valid for circular SG and WNB injections, while for the case of elliptical SG injections the energy is given as $E_{\mathrm{GW}}^{\mathrm{rot}} = (2/5) \times E_{\mathrm{GW}}^{\mathrm{iso}}$, accounting for the rotating system emission \cite{sutton-rule}.  The narrow band approximation used in this equation leads to $ \leq 6\%$ systematics in computed energy for WNB and is much lower ($\leq 3 \%$) for the SG injections.  This approximation does not hold for the GA injections, which are broadband.  The $h_\mathrm{rss}$ values for 50\% detection efficiency are used to find the typical amount of energy that needs to be radiated by the GW source in order to be detected by cWB.  These results are shown in Figure~\ref{Fig.O3EvF}.  The WNB injections for O2 are carried forward for comparison with O3.  WNB results show a factor of $2$ improvement, compatible with improvements in detector sensitivity. 

\begin{figure}[t!]
    \centering
    \scalebox{0.95}{\includegraphics[width=0.5\textwidth]{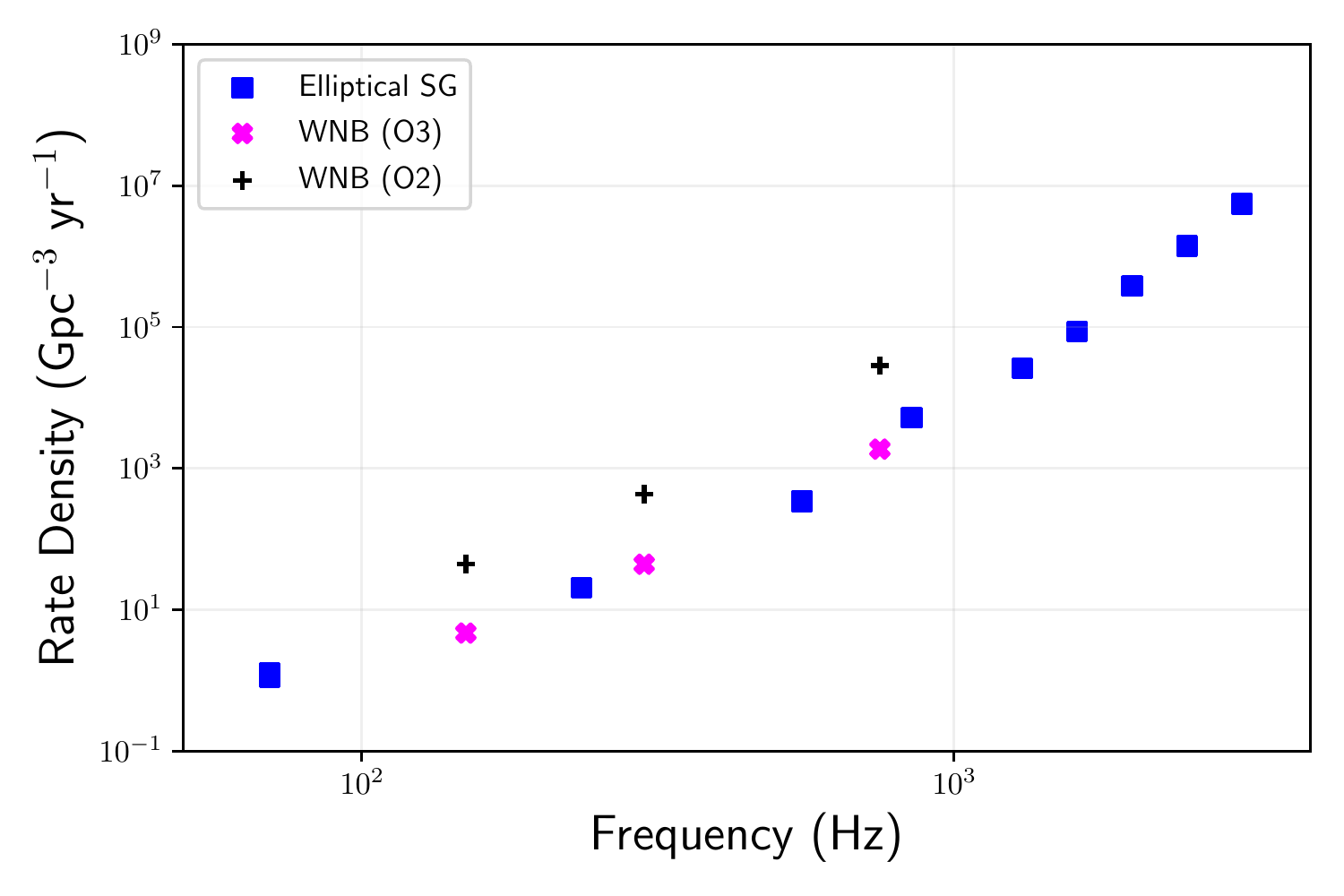}}
    \caption{
    Upper limits for the GW rate-density at 90\% confidence intervals as measured for the O3 cWB analysis using the elliptical SG and WNB waveforms are plotted assuming $1\ M_\odot c^2$ of GW energy has been emitted from the source.  For WNB waveforms the results from O2 are also plotted for comparison:  the O3 rate density upper limit is about one order of magnitude better than that achieved in O2. 
These results can be scaled to any emission energy $E_{\mathrm{GW}}$ in solar masses using the relation $\textit{rate-density} \propto E_{\mathrm{GW}}^{-3/2}$.  }
    \label{Fig.O3UpperLimits}
\end{figure}

Given that the searches do not find any GW transient sources beyond the known CBC signals, the upper limit of the rate per unit volume of non-CBC standard-candle sources~\cite{s6Burst} has been updated, as shown in Figure~\ref{Fig.O3UpperLimits}.  
These upper limits use the elliptical SG and WNB injection sets as representative morphologies for non-CBC GW bursts.
The markers represent the upper limit for rate-density at 90\% confidence~\cite{s6Burst},  calculated at an iFAR $\geq$ 100 years.  The results shown in Figure~\ref{Fig.O3UpperLimits} assume that $1\ M_\odot c^2$ of GW energy has been emitted from the source.  The upper limits can be scaled to any emission energy $E_{\mathrm{GW}}$ by using Equation~\eqref{Eq:Egw} to find that the rate-density scales as $\propto E_{\mathrm{GW}}^{-{3}/{2}}$. 
O3 results show about an order of magnitude improvement with respect to O2 for
the WNB injections.
The improvement in rate upper limits with respect to O2 is attributed to a combination of more sensitive detectors,  improved pipelines, and the longer duration of the O3 run.

\subsection{Sensitivity to CCSNe}
\label{ssec:SN_sensitivity}

Observing GWs from a CCSN would provide invaluable insight into the dynamics of these sources (e.g.,~\cite{SNGW_review}). 
Past  searches  have  looked  for  GWs  in  close spatial  and  temporal  proximity  to  electromagnetically (EM) observed CCSNe within approximately 20 Mpc \cite{targeted_SN_O1-2, targeted_SN_initialLIGO}.
It might also be possible to detect GWs from a CCSN even if its EM signatures cannot be observed, e.g.,~due to extinction along the line of sight, or in case of a failed supernova \cite{failed_SN}. Since the low-frequency unmodeled burst search presented in this paper looks for GW signals in the frequency range relevant to the majority of CCSNe, and their signal can show complex time--frequency structure, it is worthwhile to investigate the sensitivity of this search to GWs from CCSNe. The feasibility of detecting and reconstructing GWs from the next Galactic CCSN event in the upcoming observing runs are extensively studied in~\cite{Szczepanczyk:2021bka}.

Sensitivity to CCSNe is tested by analyzing waveforms from five different three-dimensional CCSN simulations. The first three represent typical CCSNe:
\begin{itemize}
 \item Model \textit{s18} \cite{Powell_2018} has a solar-metallicity non-rotating progenitor with a zero age main sequence (ZAMS) mass of 18~$M_{\odot}$. The GW emission shows the typical rise in frequency with time associated with the proto-neutron star g-mode excitation. The peak GW amplitudes occur shortly after shock revival at frequencies in the range of 800--1000~Hz.
 \item Model \textit{m20} (\textit{mesa20\_3D\_pert} from  \cite{OConnor_Couch_2018}) also corresponds to a solar-metallicity non-rotating progenitor, but it has a higher ZAMS mass of 20~$M_{\odot}$ and uses different modeling techniques. The GW emission shows the typical g-mode frequency rise, reaching $\sim$1100~Hz at the end of the simulation. Standing accretion shock instabilities (SASI, \cite{SASI1, SASI2, SASI3}) leave a subdominant imprint at frequencies of 50--100~Hz, slowly increasing in time.
 \item Model \textit{s9} \cite{Radice_2019} has a solar-metallicity non-rotating progenitor with ZAMS mass of 9~$M_\odot$. Due to its mass being in the low end of CCSNe progenitors, the density decreases rapidly outside the core and the model explodes shortly after bounce ($\sim$0.2~s). The GW signal shows the typical g-mode pattern with rising frequency and highest amplitudes within the first $\sim$0.35~s post-bounce, reaching $\sim$700~Hz.
\end{itemize}
In addition to these three models describing typical CCSNe, two simulations corresponding to more extreme CCSNe are also considered. These have higher GW amplitudes, but also lower expected rates compared to typical CCSNe (e.g.,~\cite{Janka:2012wk,Woosley:2005gy}):

\begin{itemize}
 \item Model \textit{m39} \cite{Powell_2020} describes a CCSN with a massive and rapidly rotating Wolf--Rayet star progenitor with a helium star mass of 39~$M_{\odot}$, a metallicity of 1/50 solar metallicity, and an initial surface rotation velocity of 600~km~s$^{-1}$. The rapid rotation results in larger GW amplitudes. At around 0.4~s after core bounce, the GW amplitude peaks at a frequency of $\sim$750~Hz.
 \item Model \textit{35OC} (\textit{35OC-RO} from \cite{Obergaulinger_2020}) is a simulation where the explosion is driven by strong magnetic fields and rapid rotation. The progenitor is a sub-solar metallicity star with ZAMS mass of 35~$M_{\odot}$ and equatorial surface rotation velocity of 380~km~s$^{-1}$, evolved with rotation and magnetic fields. Its high rotational energy leads to a strongly oblate shape. The waveform includes the bounce signal and oscillations above 100~Hz.
\end{itemize}
These five waveforms are chosen to represent the main physical phenomena involved and different modeling methods used.

Using each of the five CCSN models, 1000 waveforms with a uniform-in-distance distribution are generated. The maximum distance for these injections is set to $D$=[25,~5,~5,~70,~70]~kpc for the \textit{s18}, \textit{m20}, \textit{s9}, \textit{m39}, and \textit{35OC} models, respectively. All other extrinsic parameters (sky coordinates, source orientation, polarization angles) are randomized, using uniform distributions covering the full ranges of physically possible values. The sets of 1000 waveforms are repeated multiple times to cover the whole duration of the observing run.

\begin{figure}[t!]
  \centering
\includegraphics[width=1.0\columnwidth]{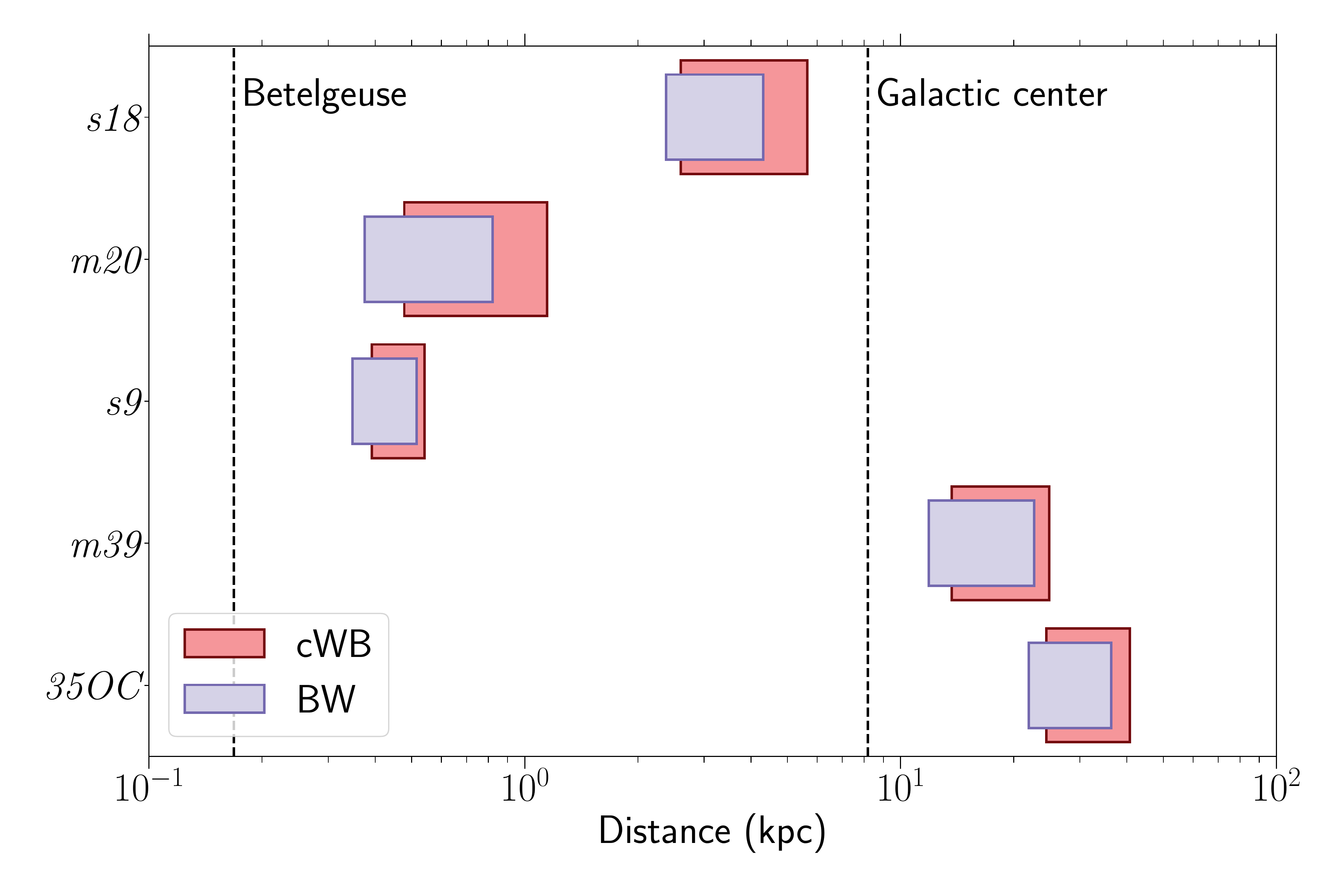}
  \caption{
  Distances at which 50\% and 10\% detection efficiencies are reached for different CCSN waveforms indicated by the left sides and right sides of rectangles, respectively. Different colors represent results from the two detection algorithms used.
  }
  \label{Fig.O3_SN_sensitivity}
\end{figure}

Results from analyzing the injections with BW and cWB are shown in Figure~\ref{Fig.O3_SN_sensitivity} as distances at which 50\% or 10\% of injected signals are detected, using the same iFAR threshold of 100~years as in Section \ref{ssec:generic_sensitivity}. The figure shows that waveforms corresponding to typical CCSNe are generally detectable only within a few kiloparsecs, while CCSNe which produce higher GW amplitudes can be observed out to tens of kiloparsecs. The cWB algorithm can detect all waveforms at similar, but slightly larger distances than BW. The largest distance at which 10\% efficiency is reached is 40.7~kpc (cWB for the \textit{35OC} model), which is smaller than the typical range of currently operating neutrino detectors (e.g.,~\cite{SK_sensitivity}). Thus any CCSN detection by the search presented in this paper would have an observable neutrino counterpart.

The same waveforms are also generated with a spatial distribution sampling the stellar mass distribution of the Milky Way, which is modeled as consisting of a bulge, a thick stellar disk and a thin stellar disk, with parameters taken from \cite{MilkyWay1} and \cite{MilkyWay2}. The overall efficiency for these injection sets is reported in Table~\ref{Tab.O3GalSN}. These represent the total fraction of simulated signals recovered, and thus are indicative of the probability that the search presented in this paper would detect a Galactic CCSN event given that the detectors were operational and under the assumption of a given CCSN model. For two typical CCSN models (\textit{m20} and \textit{s9}) the search did not detect any of the simulated signals, so an upper limit on the efficiency is quoted. This is expected, since the detector network is only sensitive to these waveforms out to $\sim$1~kpc, and the Galactic matter density model is strongly peaked around the Galactic center, so there are very few simulated signals at small distances. BW and cWB achieve low efficiencies for \textit{s18}, while they both detect a large fraction of events from the two models producing higher GW amplitudes (\textit{m39} and \textit{35OC}).

\begin{table}[!htb]
\centering
\scriptsize
\begin{tabular}{ c  c c c c c}
\hline
\hline
\textbf{Model} & \textit{s18} & \textit{m20} & \textit{s9} & \textit{m39} & \textit{35OC} \\
\hline
\textbf{cWB} & 1.2\% & <0.1\% & <0.1\% & 69.4\% & 89.8\% \\
\textbf{BW} & 0.3\% & <0.1\% & <0.1\% & 65.4\% & 89.1\% \\
\hline
\hline
\end{tabular}
\caption{The overall efficiency values with an iFAR of 100 years for each of the injected CCSN waveforms. There is a significant difference in efficiency between models of typical CCSNe and those with higher GW amplitudes. For two of the typical CCSN models (\textit{m20} and \textit{s9}) the efficiency is practically zero. This is due to the fact that these can only be detected out to $\sim$1~kpc, while the Galactic distribution provides few CCSNe at such a close distance.}\label{Tab.O3GalSN}
\end{table}

\subsection{Sensitivity to isolated neutron star emitters}

A fraction of the neutron star population is known to show transient excitations measured by EM observations.  These involve glitching pulsars and magnetars whose flaring activity include soft gamma repeaters, anomalous X-ray pulsars and giant flares.
The observed rate of such phenomena is expected to be accompanied by a larger rate of yet unobserved events.  This work focuses only on glitches,  since a dedicated search for the case of magnetar bursts is performed by a dedicated search (see \cite{Mag} for O2 results).  The two most-explored mechanisms in the literature for these neutron star excitations are starquakes and superfluid crust interactions \cite{mech_NS}. 
In the superfluid mechanism there is an interaction of internal superfluid with the solid crust of a neutron star \cite{superfluid,sf2}.  Because of superfluid vortex avalanches during the spin-up phase of pulsar glitches, the excitation of one or more families of global oscillations in the neutron star leads to a GW signal on a time scale around 40 s before the observed jump in frequency.   A search for short transient GW emission associated with oscillations of the fundamental quadrupole mode excited by a pulsar timing glitch was performed with the data from LIGO's fifth science run (S5).  No GW detection candidate was found associated with a timing glitch in the Vela Pulsar in August 2006 and a Bayesian 90\% credible upper limit of $6.3 \times 10^{-21}$ on the peak intrinsic strain amplitude of GW assuming a ring-down signal was set \cite{S6_NS}.

The precise model of the short-duration GW burst signal depends upon various considerations about the internal mechanism of the angular momentum transfer.  
The bulk emission of GW bursts is assumed to be due to f-mode excitation \cite{S6_NS,WH}.  Here it is assumed that the GW burst signal coming from the glitching neutron star is completely described by the f-mode oscillation modeled by a damped sinusoid and the optimistic scenario of the total glitch energy being converted to GW energy, $E_{\mathrm{glitch}}=E_{\mathrm{GW}}$.  The same approach was followed in previous studies \cite{S6_NS}.  

Estimates of the frequency and damping time of the neutron star fundamental quadrupole mode for various models of the equation of state (EoS) indicate that the related GW frequency is expected in the range $2 \ \mathrm{kHz} \leq \nu_{\mathrm{GW}} \leq 3 \ \mathrm{kHz}$ and the damping time is in the range of tens of milliseconds to as much as half a second \cite{WH}.  Hence, the higher frequency part of the HL all-sky search for generic bursts can survey these phenomena and motivates a dedicated astrophysical interpretation to explain the search's reach and coverage of Galactic sources.  

\begin{figure}[t!]
  \centering
\includegraphics[width=1.0\columnwidth]{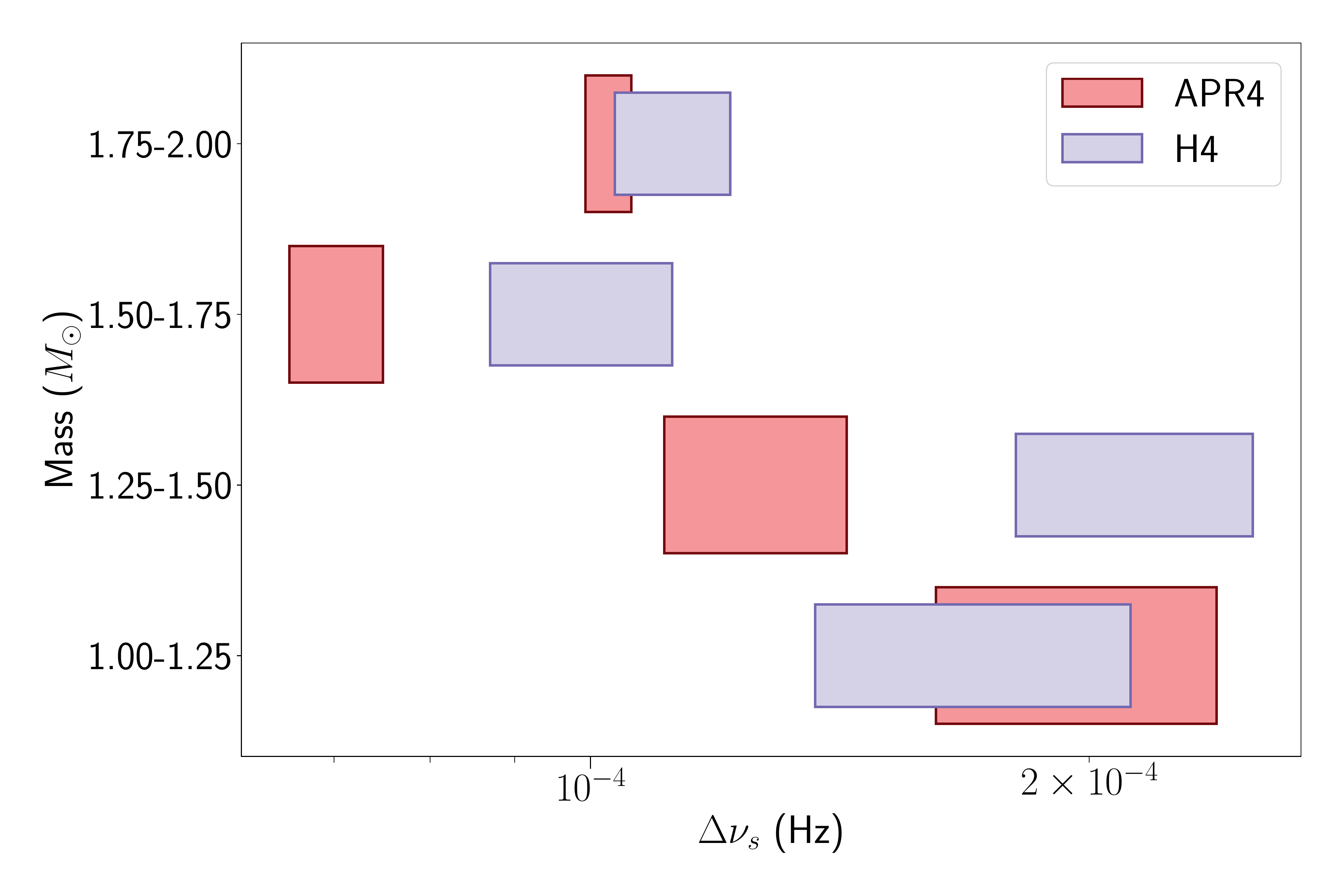}
  \caption{ Sensitivity to neutron star glitches is shown in terms of detectable glitch size by considering the Vela Pulsar as a standard candle (distance and spin of Vela) for soft (APR4) and hard (H4) EoS assuming an optimally oriented source.  For each EoS the boxes show the search sensitivity of the glitch size for 50\% detection efficiency at iFAR $\geq$ 100 years, and the spread of the box shows the variation within the mass bin. A higher-mass neutron star allows for smaller glitches to be detected. Glitch size across the parameter space for a Vela-like pulsar would need to be stronger than $ \sim $10$^{-4}$ for 50\% of the sources to be detected in O3.}
    \label{Fig.NS_sensitivity}
\end{figure}

The following discussion focuses on providing the sensitivity of the all-sky search for GWs arising from neutron star glitches.  Here the Vela Pulsar is used as a standard candle (distance of 287 pc and spin $\nu_s = 11.2$ $\text{Hz}$) to interpret the results as it is the closest known glitching pulsar \cite{vdist, vspin}.   
The signal injections are uniform in all sky directions and the source is assumed to be optimally oriented, i.e. circularly polarized.  The f-mode damped sinusoid's frequencies and damping times are related to the mass and radius of isolated neutron stars in the non-rotating limit \cite{MR-FT}.  The neutron star masses are in the range of 1--2 $M_{\odot}$ with $0.25~M_{\odot}$ bins.  The radius of the neutron star for each mass bin is determined by using two EoS,  these are APR4 (soft) \cite{apr4} and H4 (hard) \cite{h4}.  The observation of GW170817 suggests that APR4 is preferred over H4  \cite{nsradius, collab_eos_2}.  The sensitivity is determined using the $h_{\mathrm{rss}}$ values at $50\%$ detection efficiency for each mass bin and EoS.  From this the detectable glitch size $\Delta \nu_{s}$  is determined using equation 5 in \cite{WH},  assuming that the neutron star has the same distance and spin as the Vela Pulsar.  The typical $h_{\mathrm{rss}}$ at 50\% detection efficiency for an iFAR of 100 years is around $10^{-22}$~$\text{Hz}^{-1/2}$.  The sensitivities are reported in terms of glitch size as a function of mass and EoS in Figure~\ref{Fig.NS_sensitivity}.  The detectable glitch size for the O3 run is around $10^{-4}$~$\text{Hz}$, whereas the actual glitch sizes vary between $10^{-9}$--$10^{-4}$~$\text{Hz}$ \cite{gref_1,gref_2,gref_3}.  The sensitivities obtained for O3 are thus not in the range where a detection would be expected. 

\label{ssec:NS_sensitivity}

\section{Conclusion}
\label{Discussions}

This paper reports the results of a search for 
short duration GW transients of generic morphologies in O3.
The search uses minimal assumptions on the signal waveform, direction and arrival time and targets bursts with duration up to a few seconds with reconstructed central frequency from 24 Hz to 4096 Hz. 
The cWB algorithm provides results for the entire frequency range, while the BW algorithm performs a follow-up of the loudest cWB candidate events with frequencies up to 1 kHz. 
Both analyses detect GWs from CBC which have been identified by other targeted analyses for these sources.  These detections are not discussed in this paper and instead have been included in papers dedicated to CBCs \cite{GWTC-2}, or will be included in upcoming papers. 
No other significant events have been found.  The three loudest candidates remaining in the search are discussed, but their statistical significance is insufficient to exclude an instrumental origin.  Two unmodeled GW transient candidates that triggered online public alerts are also discussed, with explanations of why they do not appear in this search. 

The null result of this search allows setting of rate-density upper limits, similarly to what was done for previous observation runs \cite{s6Burst, o1allsky, allskyo2} at an inverse false alarm rate threshold of 100 years. 
The current upper limit is about one order of magnitude better than the previous O2 limit over most of the frequency bandwidth \cite{allskyo2}, mainly due to improved spectral sensitivity of the detectors and increased observation time. In addition, the typical sensitivity of this search improves by about two orders of magnitude at the lowest frequencies tested (70 Hz). The latter result stems from a combination of lower detector noise, better cleaning of data from power line sidebands, and algorithm improvements for glitch classification.  
The null results can be used to estimate sensitivity to certain classes of GW signals: CCSNe and isolated NS excitations. 
No specific tuning of the analysis is attempted, in order to preserve the general character of the search.
Five CCSN models have been tested: for the two models that produce higher GW amplitudes, the coverage of the Galaxy by this search is already good for the O3 search. However, for more typical CCSN models, the current coverage of the Galaxy is still poor. It is expected that during the next observation runs some of these, e.g., model \textit{s18}, might also achieve good Galactic coverage using GW information alone, while the distance at which CCSNe described by models producing higher GW amplitudes are detectable could reach the distance of nearby dwarf galaxies, like the Large Magellanic Cloud. 
The neutron star signals considered are f-mode emissions, modeled as damped oscillations with central frequency and damping time determined by two equations of state for the stellar mass range 1--2 $M_{\odot}$.
The sensitivities achieved by this search for generic bursts are still not sufficient to be able to detect such high-frequency transients at the energy scale of pulsar glitches from e.g., the Vela Pulsar at high confidence.
Nevertheless the outlook is promising, since the expected improvements of the GW detectors in the high-frequency band for the next observation run are quite relevant \cite{ObservingScenario}, e.g., a factor 4 and 2 in amplitude strain spectral density for Virgo and LIGO Hanford respectively. The resulting improvement on the detectable glitch size is quadratic, so near future untargeted all-sky searches for GW bursts will start probing the physical energy range observed in Vela Pulsar glitches.

\acknowledgments{}

This article has been assigned the LIGO document number P2100045.

This material is based upon work supported by NSF’s LIGO Laboratory which is a major facility fully funded by the National Science Foundation.
The authors also gratefully acknowledge 
the support of Science and Technology Facilities Council (STFC) of the
United Kingdom, the Max-Planck-Society (MPS), and the State of
Niedersachsen/Germany for support of the construction of Advanced LIGO 
and construction and operation of the GEO600 detector. 
Additional support for Advanced LIGO was provided by the Australian Research Council.
The authors gratefully acknowledge the Italian Istituto Nazionale di Fisica Nucleare (INFN),  
the French Centre National de la Recherche Scientifique (CNRS) and
the Netherlands Organization for Scientific Research, 
for the construction and operation of the Virgo detector
and the creation and support  of the EGO consortium. 
The authors also gratefully acknowledge research support from these agencies as well as by 
the Council of Scientific and Industrial Research of India, 
the Department of Science and Technology, India,
the Science \& Engineering Research Board (SERB), India,
the Ministry of Human Resource Development, India,
the Spanish Agencia Estatal de Investigaci\'on,
the Vicepresid\`encia i Conselleria d'Innovaci\'o, Recerca i Turisme and the Conselleria d'Educaci\'o i Universitat del Govern de les Illes Balears,
the Conselleria d'Innovaci\'o, Universitats, Ci\`encia i Societat Digital de la Generalitat Valenciana and
the CERCA Programme Generalitat de Catalunya, Spain,
the National Science Centre of Poland and the Foundation for Polish Science (FNP),
the Swiss National Science Foundation (SNSF),
the Russian Foundation for Basic Research, 
the Russian Science Foundation,
the European Commission,
the European Regional Development Funds (ERDF),
the Royal Society, 
the Scottish Funding Council, 
the Scottish Universities Physics Alliance, 
the Hungarian Scientific Research Fund (OTKA),
the French Lyon Institute of Origins (LIO),
the Belgian Fonds de la Recherche Scientifique (FRS-FNRS), 
Actions de Recherche Concert\'{e}es (ARC) and
Fonds Wetenschappelijk Onderzoek - Vlaanderen (FWO), Belgium,
the Paris \^{I}le-de-France Region, 
the National Research, Development and Innovation Office Hungary (NKFIH), 
the National Research Foundation of Korea,
the Natural Science and Engineering Research Council Canada,
Canadian Foundation for Innovation (CFI),
the Brazilian Ministry of Science, Technology, and Innovations,
the International Center for Theoretical Physics South American Institute for Fundamental Research (ICTP-SAIFR), 
the Research Grants Council of Hong Kong,
the National Natural Science Foundation of China (NSFC),
the Leverhulme Trust, 
the Research Corporation, 
the Ministry of Science and Technology (MOST), Taiwan,
the United States Department of Energy,
and
the Kavli Foundation.
The authors gratefully acknowledge the support of the NSF, STFC, INFN and CNRS for provision of computational resources.
This work was supported by MEXT, JSPS Leading-edge Research Infrastructure Program, JSPS Grant-in-Aid for Specially Promoted Research 26000005(Kajita 2014-2018), JSPS Grant-in-Aid for Scientific Research on Innovative Areas 2905: JP17H06358, JP17H06361 and JP17H06364, JSPS Core-to-Core Program A. Advanced Research Networks, JSPS Grant-in-Aid for Scientific Research (S) 17H06133 and 20H05639 , JSPS Grant-in-Aid for Transformative Research Areas (A) 20A203: JP20H05854, the joint research program of the Institute for Cosmic Ray Research, University of Tokyo, National Research Foundation (NRF) and Computing Infrastructure Project of KISTI-GSDC in Korea, Academia Sinica (AS), AS Grid Center (ASGC) and the Ministry of Science and Technology (MoST) in Taiwan under grants including AS-CDA-105-M06, Advanced Technology Center (ATC) of NAOJ, Mechanical Engineering Center of KEK.

We would like to thank all of the essential workers who put their health at risk during the COVID-19 pandemic, without whom we would not have been able to complete this work.

\bibliographystyle{apsrev4-1}
\bibliography{O3_allsky}

\begin{thebibliography}{81}%
\makeatletter
\providecommand \@ifxundefined [1]{%
 \@ifx{#1\undefined}
}%
\providecommand \@ifnum [1]{%
 \ifnum #1\expandafter \@firstoftwo
 \else \expandafter \@secondoftwo
 \fi
}%
\providecommand \@ifx [1]{%
 \ifx #1\expandafter \@firstoftwo
 \else \expandafter \@secondoftwo
 \fi
}%
\providecommand \natexlab [1]{#1}%
\providecommand \enquote  [1]{``#1''}%
\providecommand \bibnamefont  [1]{#1}%
\providecommand \bibfnamefont [1]{#1}%
\providecommand \citenamefont [1]{#1}%
\providecommand \href@noop [0]{\@secondoftwo}%
\providecommand \href [0]{\begingroup \@sanitize@url \@href}%
\providecommand \@href[1]{\@@startlink{#1}\@@href}%
\providecommand \@@href[1]{\endgroup#1\@@endlink}%
\providecommand \@sanitize@url [0]{\catcode `\\12\catcode `\$12\catcode
  `\&12\catcode `\#12\catcode `\^12\catcode `\_12\catcode `\%12\relax}%
\providecommand \@@startlink[1]{}%
\providecommand \@@endlink[0]{}%
\providecommand \url  [0]{\begingroup\@sanitize@url \@url }%
\providecommand \@url [1]{\endgroup\@href {#1}{\urlprefix }}%
\providecommand \urlprefix  [0]{URL }%
\providecommand \Eprint [0]{\href }%
\providecommand \doibase [0]{http://dx.doi.org/}%
\providecommand \selectlanguage [0]{\@gobble}%
\providecommand \bibinfo  [0]{\@secondoftwo}%
\providecommand \bibfield  [0]{\@secondoftwo}%
\providecommand \translation [1]{[#1]}%
\providecommand \BibitemOpen [0]{}%
\providecommand \bibitemStop [0]{}%
\providecommand \bibitemNoStop [0]{.\EOS\space}%
\providecommand \EOS [0]{\spacefactor3000\relax}%
\providecommand \BibitemShut  [1]{\csname bibitem#1\endcsname}%
\let\auto@bib@innerbib\@empty
%</preamble>
\bibitem [{\citenamefont {Aasi}\ \emph {et~al.}(2015)\citenamefont {Aasi} \emph
  {et~al.}}]{LIGO-det}%
  \BibitemOpen
  \bibfield  {author} {\bibinfo {author} {\bibfnamefont {J.}~\bibnamefont
  {Aasi}} \emph {et~al.} (\bibinfo {collaboration} {LIGO Scientific
  Collaboration}),\ }\href {\doibase 10.1088/0264-9381/32/7/074001} {\bibfield
  {journal} {\bibinfo  {journal} {Class. Quant. Grav.}\ }\textbf {\bibinfo
  {volume} {32}},\ \bibinfo {pages} {074001} (\bibinfo {year}
  {2015})}\BibitemShut {NoStop}%
\bibitem [{\citenamefont {Acernese}\ \emph {et~al.}(2015)\citenamefont
  {Acernese} \emph {et~al.}}]{Virgo-det}%
  \BibitemOpen
  \bibfield  {author} {\bibinfo {author} {\bibfnamefont {F.}~\bibnamefont
  {Acernese}} \emph {et~al.} (\bibinfo {collaboration} {Virgo Collaboration}),\
  }\href {\doibase 10.1088/0264-9381/32/2/024001} {\bibfield  {journal}
  {\bibinfo  {journal} {Class. Quant. Grav.}\ }\textbf {\bibinfo {volume}
  {32}},\ \bibinfo {pages} {024001} (\bibinfo {year} {2015})}\BibitemShut
  {NoStop}%
\bibitem [{\citenamefont {Abbott}\ \emph {et~al.}(2021)\citenamefont {Abbott}
  \emph {et~al.}}]{GWTC-2}%
  \BibitemOpen
  \bibfield  {author} {\bibinfo {author} {\bibfnamefont {R.}~\bibnamefont
  {Abbott}} \emph {et~al.} (\bibinfo {collaboration} {LIGO Scientific
  Collaboration and Virgo Collaboration}),\ }\href
  {https://journals.aps.org/prx/abstract/10.1103/PhysRevX.11.021053} {\bibfield
   {journal} {\bibinfo  {journal} {Phys. Rev. X}\ }\textbf {\bibinfo {volume}
  {11}},\ \bibinfo {pages} {021053} (\bibinfo {year} {2021})}\BibitemShut
  {NoStop}%
\bibitem [{\citenamefont {{Abbott}}\ \emph {et~al.}(2021)\citenamefont
  {{Abbott}} \emph {et~al.}}]{O3IMBH}%
  \BibitemOpen
  \bibfield  {author} {\bibinfo {author} {\bibfnamefont {R.}~\bibnamefont
  {{Abbott}}} \emph {et~al.} (\bibinfo {collaboration} {LIGO Scientific
  Collaboration, Virgo Collaboration and KAGRA Collaboration}),\ }\href@noop {}
  {\bibfield  {journal} {\bibinfo  {journal} {{arXiv e-prints}}\ } (\bibinfo
  {year} {2021})},\ \Eprint {http://arxiv.org/abs/2105.15120} {arXiv:2105.15120
  [astro-ph]} \BibitemShut {NoStop}%
\bibitem [{\citenamefont {Abbott}\ \emph
  {et~al.}(2021{\natexlab{a}})\citenamefont {Abbott} \emph {et~al.}}]{NSBH}%
  \BibitemOpen
  \bibfield  {author} {\bibinfo {author} {\bibfnamefont {R.}~\bibnamefont
  {Abbott}} \emph {et~al.} (\bibinfo {collaboration} {LIGO Scientific
  Collaboration and Virgo Collaboration}),\ }\href@noop {} {\bibfield
  {journal} {\bibinfo  {journal} {in preparation}\ } (\bibinfo {year}
  {2021}{\natexlab{a}})}\BibitemShut {NoStop}%
\bibitem [{\citenamefont {Nitz}\ \emph {et~al.}(2021)\citenamefont {Nitz},
  \citenamefont {Capano}, \citenamefont {Kumar}, \citenamefont {Wang},
  \citenamefont {Kastha}, \citenamefont {Sch{\"a}fer}, \citenamefont
  {Dhurkunde},\ and\ \citenamefont {Cabero}}]{3-OGC}%
  \BibitemOpen
  \bibfield  {author} {\bibinfo {author} {\bibfnamefont {A.~H.}\ \bibnamefont
  {Nitz}}, \bibinfo {author} {\bibfnamefont {C.~D.}\ \bibnamefont {Capano}},
  \bibinfo {author} {\bibfnamefont {S.}~\bibnamefont {Kumar}}, \bibinfo
  {author} {\bibfnamefont {Y.-F.}\ \bibnamefont {Wang}}, \bibinfo {author}
  {\bibfnamefont {S.}~\bibnamefont {Kastha}}, \bibinfo {author} {\bibfnamefont
  {M.}~\bibnamefont {Sch{\"a}fer}}, \bibinfo {author} {\bibfnamefont
  {R.}~\bibnamefont {Dhurkunde}}, \ and\ \bibinfo {author} {\bibfnamefont
  {M.}~\bibnamefont {Cabero}},\ }\href@noop {} {\bibfield  {journal} {\bibinfo
  {journal} {arXiv e-prints}\ } (\bibinfo {year} {2021})},\ \Eprint
  {http://arxiv.org/abs/2105.09151} {arXiv:2105.09151 [astro-ph.HE]}
  \BibitemShut {NoStop}%
\bibitem [{\citenamefont {{Abbott}}\ \emph {et~al.}(2020)\citenamefont
  {{Abbott}} \emph {et~al.}}]{targeted_SN_O1-2}%
  \BibitemOpen
  \bibfield  {author} {\bibinfo {author} {\bibfnamefont {B.~P.}\ \bibnamefont
  {{Abbott}}} \emph {et~al.} (\bibinfo {collaboration} {LIGO Scientific
  Collaboration and Virgo Collaboration}),\ }\href {\doibase
  10.1103/PhysRevD.101.084002} {\bibfield  {journal} {\bibinfo  {journal}
  {\prd}\ }\textbf {\bibinfo {volume} {101}},\ \bibinfo {eid} {084002}
  (\bibinfo {year} {2020})}\BibitemShut {NoStop}%
\bibitem [{\citenamefont {Abbott}\ \emph
  {et~al.}(2019{\natexlab{a}})\citenamefont {Abbott} \emph
  {et~al.}}]{O2magnetarbursts}%
  \BibitemOpen
  \bibfield  {author} {\bibinfo {author} {\bibfnamefont {B.~P.}\ \bibnamefont
  {Abbott}} \emph {et~al.} (\bibinfo {collaboration} {LIGO Scientific
  Collaboration and Virgo Collaboration}),\ }\href {\doibase
  10.3847/1538-4357/ab0e15} {\bibfield  {journal} {\bibinfo  {journal}
  {Astrophys. J.}\ }\textbf {\bibinfo {volume} {874}},\ \bibinfo {pages} {163}
  (\bibinfo {year} {2019}{\natexlab{a}})}\BibitemShut {NoStop}%
\bibitem [{\citenamefont {Abadie}\ \emph {et~al.}(2011)\citenamefont {Abadie}
  \emph {et~al.}}]{S6_NS}%
  \BibitemOpen
  \bibfield  {author} {\bibinfo {author} {\bibfnamefont {J.}~\bibnamefont
  {Abadie}} \emph {et~al.} (\bibinfo {collaboration} {LIGO Scientific
  Collaboration}),\ }\href
  {https://journals.aps.org/prd/abstract/10.1103/PhysRevD.83.042001} {\bibfield
   {journal} {\bibinfo  {journal} {Phys. Rev. D}\ }\textbf {\bibinfo {volume}
  {83}},\ \bibinfo {pages} {042001} (\bibinfo {year} {2011})}\BibitemShut
  {NoStop}%
\bibitem [{\citenamefont {Ebersold}\ and\ \citenamefont
  {Tiwari}(2020)}]{o2_mem}%
  \BibitemOpen
  \bibfield  {author} {\bibinfo {author} {\bibfnamefont {M.}~\bibnamefont
  {Ebersold}}\ and\ \bibinfo {author} {\bibfnamefont {S.}~\bibnamefont
  {Tiwari}},\ }\href {\doibase 10.1103/PhysRevD.101.104041} {\bibfield
  {journal} {\bibinfo  {journal} {Phys. Rev. D}\ }\textbf {\bibinfo {volume}
  {101}},\ \bibinfo {pages} {104041} (\bibinfo {year} {2020})}\BibitemShut
  {NoStop}%
\bibitem [{\citenamefont {Abbott}\ \emph
  {et~al.}(2021{\natexlab{b}})\citenamefont {Abbott} \emph
  {et~al.}}]{O3cosmicstring}%
  \BibitemOpen
  \bibfield  {author} {\bibinfo {author} {\bibfnamefont {R.}~\bibnamefont
  {Abbott}} \emph {et~al.} (\bibinfo {collaboration} {LIGO Scientific
  Collaboration, Virgo Collaboration and KAGRA Collaboration}),\ }\href
  {https://journals.aps.org/prl/abstract/10.1103/PhysRevLett.126.241102}
  {\bibfield  {journal} {\bibinfo  {journal} {Phys. Rev. Lett.}\ }\textbf
  {\bibinfo {volume} {126}},\ \bibinfo {pages} {241102} (\bibinfo {year}
  {2021}{\natexlab{b}})}\BibitemShut {NoStop}%
\bibitem [{\citenamefont {Abbott}\ \emph
  {et~al.}(2021{\natexlab{c}})\citenamefont {Abbott} \emph
  {et~al.}}]{O3alensing}%
  \BibitemOpen
  \bibfield  {author} {\bibinfo {author} {\bibfnamefont {R.}~\bibnamefont
  {Abbott}} \emph {et~al.} (\bibinfo {collaboration} {LIGO Scientific
  Collaboration and Virgo Collaboration}),\ }\href@noop {} {\bibfield
  {journal} {\bibinfo  {journal} {arXiv e-prints}\ } (\bibinfo {year}
  {2021}{\natexlab{c}})},\ \Eprint {http://arxiv.org/abs/2105.06384}
  {arXiv:2105.06384 [gr-qc]} \BibitemShut {NoStop}%
\bibitem [{\citenamefont {Abbott}\ \emph
  {et~al.}(2019{\natexlab{b}})\citenamefont {Abbott} \emph
  {et~al.}}]{allskyo2}%
  \BibitemOpen
  \bibfield  {author} {\bibinfo {author} {\bibfnamefont {B.~P.}\ \bibnamefont
  {Abbott}} \emph {et~al.} (\bibinfo {collaboration} {LIGO Scientific
  Collaboration, Virgo Collaboration}),\ }\href
  {http://dx.doi.org/10.1103/PhysRevD.100.024017} {\bibfield  {journal}
  {\bibinfo  {journal} {Phys. Rev. D}\ }\textbf {\bibinfo {volume} {100}},\
  \bibinfo {pages} {024017} (\bibinfo {year} {2019}{\natexlab{b}})}\BibitemShut
  {NoStop}%
\bibitem [{\citenamefont {Abbott}\ \emph
  {et~al.}(2019{\natexlab{c}})\citenamefont {Abbott} \emph
  {et~al.}}]{allskyo2-long}%
  \BibitemOpen
  \bibfield  {author} {\bibinfo {author} {\bibfnamefont {B.~P.}\ \bibnamefont
  {Abbott}} \emph {et~al.} (\bibinfo {collaboration} {LIGO Scientific
  Collaboration and Virgo Collaboration}),\ }\href {\doibase
  10.1103/PhysRevD.99.104033} {\bibfield  {journal} {\bibinfo  {journal} {Phys.
  Rev. D}\ }\textbf {\bibinfo {volume} {99}},\ \bibinfo {pages} {104033}
  (\bibinfo {year} {2019}{\natexlab{c}})}\BibitemShut {NoStop}%
\bibitem [{\citenamefont {Abbott}\ \emph {et~al.}(2017)\citenamefont {Abbott}
  \emph {et~al.}}]{o1allsky}%
  \BibitemOpen
  \bibfield  {author} {\bibinfo {author} {\bibfnamefont {B.~P.}\ \bibnamefont
  {Abbott}} \emph {et~al.} (\bibinfo {collaboration} {LIGO Scientific
  Collaboration and Virgo Collaboration}),\ }\href {\doibase
  10.1103/PhysRevD.95.042003} {\bibfield  {journal} {\bibinfo  {journal} {Phys.
  Rev. D}\ }\textbf {\bibinfo {volume} {95}},\ \bibinfo {pages} {042003}
  (\bibinfo {year} {2017})}\BibitemShut {NoStop}%
\bibitem [{\citenamefont {{Sun}}\ \emph {et~al.}(2020)\citenamefont {{Sun}},
  \citenamefont {{Goetz}}, \citenamefont {{Kissel}}, \citenamefont
  {{Betzwieser}}, \citenamefont {{Karki}}, \citenamefont {{Viets}},
  \citenamefont {{Wade}}, \citenamefont {{Bhattacharjee}}, \citenamefont
  {{Bossilkov}}, \citenamefont {{Covas}}, \citenamefont {{Datrier}},
  \citenamefont {{Gray}}, \citenamefont {{Kandhasamy}}, \citenamefont
  {{Lecoeuche}}, \citenamefont {{Mendell}}, \citenamefont {{Mistry}},
  \citenamefont {{Payne}}, \citenamefont {{Savage}}, \citenamefont
  {{Weinstein}}, \citenamefont {{Aston}}, \citenamefont {{Buikema}},
  \citenamefont {{Cahillane}}, \citenamefont {{Driggers}}, \citenamefont
  {{Dwyer}}, \citenamefont {{Kumar}},\ and\ \citenamefont
  {{Urban}}}]{Sun_et_al_2020}%
  \BibitemOpen
  \bibfield  {author} {\bibinfo {author} {\bibfnamefont {L.}~\bibnamefont
  {{Sun}}}, \bibinfo {author} {\bibfnamefont {E.}~\bibnamefont {{Goetz}}},
  \bibinfo {author} {\bibfnamefont {J.~S.}\ \bibnamefont {{Kissel}}}, \bibinfo
  {author} {\bibfnamefont {J.}~\bibnamefont {{Betzwieser}}}, \bibinfo {author}
  {\bibfnamefont {S.}~\bibnamefont {{Karki}}}, \bibinfo {author} {\bibfnamefont
  {A.}~\bibnamefont {{Viets}}}, \bibinfo {author} {\bibfnamefont
  {M.}~\bibnamefont {{Wade}}}, \bibinfo {author} {\bibfnamefont
  {D.}~\bibnamefont {{Bhattacharjee}}}, \bibinfo {author} {\bibfnamefont
  {V.}~\bibnamefont {{Bossilkov}}}, \bibinfo {author} {\bibfnamefont {P.~B.}\
  \bibnamefont {{Covas}}}, \bibinfo {author} {\bibfnamefont {L.~E.~H.}\
  \bibnamefont {{Datrier}}}, \bibinfo {author} {\bibfnamefont {R.}~\bibnamefont
  {{Gray}}}, \bibinfo {author} {\bibfnamefont {S.}~\bibnamefont
  {{Kandhasamy}}}, \bibinfo {author} {\bibfnamefont {Y.~K.}\ \bibnamefont
  {{Lecoeuche}}}, \bibinfo {author} {\bibfnamefont {G.}~\bibnamefont
  {{Mendell}}}, \bibinfo {author} {\bibfnamefont {T.}~\bibnamefont {{Mistry}}},
  \bibinfo {author} {\bibfnamefont {E.}~\bibnamefont {{Payne}}}, \bibinfo
  {author} {\bibfnamefont {R.~L.}\ \bibnamefont {{Savage}}}, \bibinfo {author}
  {\bibfnamefont {A.~J.}\ \bibnamefont {{Weinstein}}}, \bibinfo {author}
  {\bibfnamefont {S.}~\bibnamefont {{Aston}}}, \bibinfo {author} {\bibfnamefont
  {A.}~\bibnamefont {{Buikema}}}, \bibinfo {author} {\bibfnamefont
  {C.}~\bibnamefont {{Cahillane}}}, \bibinfo {author} {\bibfnamefont {J.~C.}\
  \bibnamefont {{Driggers}}}, \bibinfo {author} {\bibfnamefont {S.~E.}\
  \bibnamefont {{Dwyer}}}, \bibinfo {author} {\bibfnamefont {R.}~\bibnamefont
  {{Kumar}}}, \ and\ \bibinfo {author} {\bibfnamefont {A.}~\bibnamefont
  {{Urban}}},\ }\href {\doibase 10.1088/1361-6382/abb14e} {\bibfield  {journal}
  {\bibinfo  {journal} {Class. Quant. Grav.}\ }\textbf {\bibinfo {volume}
  {37}},\ \bibinfo {eid} {225008} (\bibinfo {year} {2020})}\BibitemShut
  {NoStop}%
\bibitem [{\citenamefont {{Sun}}\ \emph {et~al.}(2021)\citenamefont {{Sun}},
  \citenamefont {Kissel}, \citenamefont {Goetz},\ and\ \citenamefont
  {Betzweiser}}]{O3b_LIGO_calib}%
  \BibitemOpen
  \bibfield  {author} {\bibinfo {author} {\bibfnamefont {L.}~\bibnamefont
  {{Sun}}}, \bibinfo {author} {\bibfnamefont {J.}~\bibnamefont {Kissel}},
  \bibinfo {author} {\bibfnamefont {E.}~\bibnamefont {Goetz}}, \ and\ \bibinfo
  {author} {\bibfnamefont {J.}~\bibnamefont {Betzweiser}},\ }\href
  {https://dcc.ligo.org/LIGO-T2000584/public} {\emph {\bibinfo {title}
  {{Characterization of systematic error in Advanced LIGO calibration in the
  second half of O3 (O3B)}}}},\ \bibinfo {type} {Tech. Rep.}\ \bibinfo {number}
  {LIGO-T2000584}\ (\bibinfo  {institution} {LIGO},\ \bibinfo {year}
  {2021})\BibitemShut {NoStop}%
\bibitem [{\citenamefont {Estevez}\ \emph {et~al.}(2019)\citenamefont
  {Estevez}, \citenamefont {Mours}, \citenamefont {Rolland},\ and\
  \citenamefont {D.Verkindt}}]{O3a_Virgo_calib}%
  \BibitemOpen
  \bibfield  {author} {\bibinfo {author} {\bibfnamefont {D.}~\bibnamefont
  {Estevez}}, \bibinfo {author} {\bibfnamefont {B.}~\bibnamefont {Mours}},
  \bibinfo {author} {\bibfnamefont {L.}~\bibnamefont {Rolland}}, \ and\
  \bibinfo {author} {\bibnamefont {D.Verkindt}},\ }\href
  {https://tds.ego-gw.it/ql/?c=14486} {\emph {\bibinfo {title} {{Online h(t)
  reconstruction for Virgo O3 data: start of O3}}}},\ \bibinfo {type} {Tech.
  Rep.}\ \bibinfo {number} {VIR-0652B-19}\ (\bibinfo  {institution} {Virgo},\
  \bibinfo {year} {2019})\BibitemShut {NoStop}%
\bibitem [{\citenamefont {{Davis}}\ \emph {et~al.}(2021)\citenamefont {{Davis}}
  \emph {et~al.}}]{O2-O3-detchar}%
  \BibitemOpen
  \bibfield  {author} {\bibinfo {author} {\bibfnamefont {D.}~\bibnamefont
  {{Davis}}} \emph {et~al.},\ }\href
  {https://iopscience.iop.org/article/10.1088/1361-6382/abfd85} {\bibfield
  {journal} {\bibinfo  {journal} {Class. Quant. Grav.}\ }\textbf {\bibinfo
  {volume} {38}},\ \bibinfo {pages} {135014} (\bibinfo {year}
  {2021})}\BibitemShut {NoStop}%
\bibitem [{pub()}]{publicalertsuserguide}%
  \BibitemOpen
  \href@noop {} {}\bibinfo {howpublished}
  {\url{https://emfollow.docs.ligo.org/userguide/}}\BibitemShut {NoStop}%
\bibitem [{\citenamefont {Rolland}\ \emph {et~al.}(2020)\citenamefont
  {Rolland}, \citenamefont {Estevez}, \citenamefont {Lagabbe}, \citenamefont
  {Mours}, \citenamefont {Seglar-Arroyo}, \citenamefont {Pradier},\ and\
  \citenamefont {D.}}]{Rolland:2020vna}%
  \BibitemOpen
  \bibfield  {author} {\bibinfo {author} {\bibfnamefont {L.}~\bibnamefont
  {Rolland}}, \bibinfo {author} {\bibfnamefont {D.}~\bibnamefont {Estevez}},
  \bibinfo {author} {\bibfnamefont {P.}~\bibnamefont {Lagabbe}}, \bibinfo
  {author} {\bibfnamefont {B.}~\bibnamefont {Mours}}, \bibinfo {author}
  {\bibfnamefont {M.}~\bibnamefont {Seglar-Arroyo}}, \bibinfo {author}
  {\bibfnamefont {T.}~\bibnamefont {Pradier}}, \ and\ \bibinfo {author}
  {\bibfnamefont {V.}~\bibnamefont {D.}},\ }\href
  {https://tds.virgo-gw.eu/ql/?c=15813} {\emph {\bibinfo {title} {{Update on
  h(t) uncertainties during O3}}}},\ \bibinfo {type} {Tech. Rep.}\ \bibinfo
  {number} {VIR-0688A-20}\ (\bibinfo  {institution} {{Virgo}},\ \bibinfo {year}
  {2020})\BibitemShut {NoStop}%
\bibitem [{\citenamefont {Abbott}\ \emph
  {et~al.}(2021{\natexlab{d}})\citenamefont {Abbott} \emph {et~al.}}]{GWOSC}%
  \BibitemOpen
  \bibfield  {author} {\bibinfo {author} {\bibfnamefont {R.}~\bibnamefont
  {Abbott}} \emph {et~al.} (\bibinfo {collaboration} {LIGO Scientific
  Collaboration and Virgo Collaboration}),\ }\href {\doibase
  10.1016/j.softx.2021.100658} {\bibfield  {journal} {\bibinfo  {journal}
  {SoftwareX}\ }\textbf {\bibinfo {volume} {13}},\ \bibinfo {pages} {100658}
  (\bibinfo {year} {2021}{\natexlab{d}})}\BibitemShut {NoStop}%
\bibitem [{\citenamefont {{LIGO Scientific Collaboration And Virgo
  Collaboration}}(2021)}]{GWOSC_O3a}%
  \BibitemOpen
  \bibfield  {author} {\bibinfo {author} {\bibnamefont {{LIGO Scientific
  Collaboration And Virgo Collaboration}}},\ }\href {\doibase
  10.7935/nfnt-hm34} {\enquote {\bibinfo {title} {{LIGO} {V}irgo strain data
  from observing run {O}3a},}\ }\bibinfo {howpublished}
  {\url{https://www.gw-openscience.org/O3/O3a}} (\bibinfo {year}
  {2021})\BibitemShut {NoStop}%
\bibitem [{\citenamefont {Buikema}\ \emph {et~al.}(2020)\citenamefont {Buikema}
  \emph {et~al.}}]{O3performance}%
  \BibitemOpen
  \bibfield  {author} {\bibinfo {author} {\bibfnamefont {A.}~\bibnamefont
  {Buikema}} \emph {et~al.},\ }\href {\doibase 10.1103/PhysRevD.102.062003}
  {\bibfield  {journal} {\bibinfo  {journal} {Phys. Rev. D}\ }\textbf {\bibinfo
  {volume} {102}},\ \bibinfo {pages} {062003} (\bibinfo {year}
  {2020})}\BibitemShut {NoStop}%
\bibitem [{\citenamefont {Abbott}\ \emph {et~al.}(2016)\citenamefont {Abbott}
  \emph {et~al.}}]{Abbott_2016}%
  \BibitemOpen
  \bibfield  {author} {\bibinfo {author} {\bibfnamefont {B.~P.}\ \bibnamefont
  {Abbott}} \emph {et~al.},\ }\href {\doibase 10.1088/0264-9381/33/13/134001}
  {\bibfield  {journal} {\bibinfo  {journal} {Class. Quant. Grav.}\ }\textbf
  {\bibinfo {volume} {33}},\ \bibinfo {pages} {134001} (\bibinfo {year}
  {2016})}\BibitemShut {NoStop}%
\bibitem [{\citenamefont {Nguyen}\ \emph {et~al.}(2021)\citenamefont {Nguyen}
  \emph {et~al.}}]{envnoise}%
  \BibitemOpen
  \bibfield  {author} {\bibinfo {author} {\bibfnamefont {P.}~\bibnamefont
  {Nguyen}} \emph {et~al.},\ }\href@noop {} {\bibfield  {journal} {\bibinfo
  {journal} {arXiv e-prints}\ } (\bibinfo {year} {2021})},\ \Eprint
  {http://arxiv.org/abs/2101.09935} {arXiv:2101.09935 [astro-ph.IM]}
  \BibitemShut {NoStop}%
\bibitem [{\citenamefont {Effler}\ \emph {et~al.}(2015)\citenamefont {Effler},
  \citenamefont {Schofield}, \citenamefont {Frolov}, \citenamefont
  {Gonz{\'{a}}lez}, \citenamefont {Kawabe}, \citenamefont {Smith},
  \citenamefont {Birch},\ and\ \citenamefont {McCarthy}}]{PEM}%
  \BibitemOpen
  \bibfield  {author} {\bibinfo {author} {\bibfnamefont {A.}~\bibnamefont
  {Effler}}, \bibinfo {author} {\bibfnamefont {R.~M.~S.}\ \bibnamefont
  {Schofield}}, \bibinfo {author} {\bibfnamefont {V.~V.}\ \bibnamefont
  {Frolov}}, \bibinfo {author} {\bibfnamefont {G.}~\bibnamefont
  {Gonz{\'{a}}lez}}, \bibinfo {author} {\bibfnamefont {K.}~\bibnamefont
  {Kawabe}}, \bibinfo {author} {\bibfnamefont {J.~R.}\ \bibnamefont {Smith}},
  \bibinfo {author} {\bibfnamefont {J.}~\bibnamefont {Birch}}, \ and\ \bibinfo
  {author} {\bibfnamefont {R.}~\bibnamefont {McCarthy}},\ }\href {\doibase
  10.1088/0264-9381/32/3/035017} {\bibfield  {journal} {\bibinfo  {journal}
  {Class. Quant. Grav.}\ }\textbf {\bibinfo {volume} {32}},\ \bibinfo {pages}
  {035017} (\bibinfo {year} {2015})}\BibitemShut {NoStop}%
\bibitem [{\citenamefont {Abbott}\ \emph
  {et~al.}(2018{\natexlab{a}})\citenamefont {Abbott} \emph
  {et~al.}}]{dqmitigation}%
  \BibitemOpen
  \bibfield  {author} {\bibinfo {author} {\bibfnamefont {B.~P.}\ \bibnamefont
  {Abbott}} \emph {et~al.} (\bibinfo {collaboration} {LIGO Scientific
  Collaboration and Virgo Collaboration}),\ }\href
  {https://iopscience.iop.org/article/10.1088/1361-6382/aaaafa} {\bibfield
  {journal} {\bibinfo  {journal} {Class. Quant. Grav.}\ }\textbf {\bibinfo
  {volume} {35}},\ \bibinfo {pages} {065010} (\bibinfo {year}
  {2018}{\natexlab{a}})}\BibitemShut {NoStop}%
\bibitem [{\citenamefont {Smith}\ \emph {et~al.}(2011)\citenamefont {Smith},
  \citenamefont {Abbott}, \citenamefont {Hirose}, \citenamefont {Leroy},
  \citenamefont {MacLeod}, \citenamefont {McIver}, \citenamefont {Saulson},\
  and\ \citenamefont {Shawhan}}]{Smith_2011}%
  \BibitemOpen
  \bibfield  {author} {\bibinfo {author} {\bibfnamefont {J.~R.}\ \bibnamefont
  {Smith}}, \bibinfo {author} {\bibfnamefont {T.}~\bibnamefont {Abbott}},
  \bibinfo {author} {\bibfnamefont {E.}~\bibnamefont {Hirose}}, \bibinfo
  {author} {\bibfnamefont {N.}~\bibnamefont {Leroy}}, \bibinfo {author}
  {\bibfnamefont {D.}~\bibnamefont {MacLeod}}, \bibinfo {author} {\bibfnamefont
  {J.}~\bibnamefont {McIver}}, \bibinfo {author} {\bibfnamefont
  {P.}~\bibnamefont {Saulson}}, \ and\ \bibinfo {author} {\bibfnamefont
  {P.}~\bibnamefont {Shawhan}},\ }\href {\doibase
  10.1088/0264-9381/28/23/235005} {\bibfield  {journal} {\bibinfo  {journal}
  {Class. Quant. Grav.}\ }\textbf {\bibinfo {volume} {28}},\ \bibinfo {pages}
  {235005} (\bibinfo {year} {2011})}\BibitemShut {NoStop}%
\bibitem [{\citenamefont {Davis}\ \emph {et~al.}(2021)\citenamefont {Davis},
  \citenamefont {Hughey}, \citenamefont {Massinger}, \citenamefont {Nuttall},
  \citenamefont {Stuver},\ and\ \citenamefont {Zweizig}}]{o3dataflags}%
  \BibitemOpen
  \bibfield  {author} {\bibinfo {author} {\bibfnamefont {D.}~\bibnamefont
  {Davis}}, \bibinfo {author} {\bibfnamefont {B.}~\bibnamefont {Hughey}},
  \bibinfo {author} {\bibfnamefont {T.}~\bibnamefont {Massinger}}, \bibinfo
  {author} {\bibfnamefont {L.}~\bibnamefont {Nuttall}}, \bibinfo {author}
  {\bibfnamefont {A.}~\bibnamefont {Stuver}}, \ and\ \bibinfo {author}
  {\bibfnamefont {J.}~\bibnamefont {Zweizig}},\ }\href
  {https://dcc.ligo.org/LIGO-T2100045/public} {\emph {\bibinfo {title} {Data
  Quality Vetoes Applied to the Analysis of LIGO Data from the Third Observing
  Run}}},\ \bibinfo {type} {Tech. Rep.}\ \bibinfo {number} {LIGO-T2100045}\
  (\bibinfo  {institution} {LVK},\ \bibinfo {year} {2021})\BibitemShut
  {NoStop}%
\bibitem [{\citenamefont {Soni}\ \emph
  {et~al.}(2021{\natexlab{a}})\citenamefont {Soni} \emph {et~al.}}]{Soni_2021}%
  \BibitemOpen
  \bibfield  {author} {\bibinfo {author} {\bibfnamefont {S.}~\bibnamefont
  {Soni}} \emph {et~al.},\ }\href {\doibase 10.1088/1361-6382/abc906}
  {\bibfield  {journal} {\bibinfo  {journal} {Class. Quant. Grav.}\ }\textbf
  {\bibinfo {volume} {38}},\ \bibinfo {pages} {025016} (\bibinfo {year}
  {2021}{\natexlab{a}})}\BibitemShut {NoStop}%
\bibitem [{\citenamefont {Soni}\ \emph
  {et~al.}(2021{\natexlab{b}})\citenamefont {Soni}, \citenamefont {Berry},
  \citenamefont {Coughlin}, \citenamefont {Harandi}, \citenamefont {Jackson},
  \citenamefont {Crowston}, \citenamefont {{\O}sterlund}, \citenamefont
  {Patane}, \citenamefont {Katsaggelos}, \citenamefont {Trouille},
  \citenamefont {Baranowski}, \citenamefont {Domainko}, \citenamefont
  {Kaminski}, \citenamefont {Rodriguez}, \citenamefont {Marciniak},
  \citenamefont {Nauta}, \citenamefont {Niklasch}, \citenamefont {Rote},
  \citenamefont {T\'egl\'as}, \citenamefont {Unsworth},\ and\ \citenamefont
  {Zhang}}]{Soni_2021b}%
  \BibitemOpen
  \bibfield  {author} {\bibinfo {author} {\bibfnamefont {S.}~\bibnamefont
  {Soni}}, \bibinfo {author} {\bibfnamefont {C.~P.~L.}\ \bibnamefont {Berry}},
  \bibinfo {author} {\bibfnamefont {S.~B.}\ \bibnamefont {Coughlin}}, \bibinfo
  {author} {\bibfnamefont {M.}~\bibnamefont {Harandi}}, \bibinfo {author}
  {\bibfnamefont {C.~B.}\ \bibnamefont {Jackson}}, \bibinfo {author}
  {\bibfnamefont {K.}~\bibnamefont {Crowston}}, \bibinfo {author}
  {\bibfnamefont {C.}~\bibnamefont {{\O}sterlund}}, \bibinfo {author}
  {\bibfnamefont {O.}~\bibnamefont {Patane}}, \bibinfo {author} {\bibfnamefont
  {A.~K.}\ \bibnamefont {Katsaggelos}}, \bibinfo {author} {\bibfnamefont
  {L.}~\bibnamefont {Trouille}}, \bibinfo {author} {\bibfnamefont {V.-G.}\
  \bibnamefont {Baranowski}}, \bibinfo {author} {\bibfnamefont {W.~F.}\
  \bibnamefont {Domainko}}, \bibinfo {author} {\bibfnamefont {K.}~\bibnamefont
  {Kaminski}}, \bibinfo {author} {\bibfnamefont {M.~A.~L.}\ \bibnamefont
  {Rodriguez}}, \bibinfo {author} {\bibfnamefont {U.}~\bibnamefont
  {Marciniak}}, \bibinfo {author} {\bibfnamefont {P.}~\bibnamefont {Nauta}},
  \bibinfo {author} {\bibfnamefont {G.}~\bibnamefont {Niklasch}}, \bibinfo
  {author} {\bibfnamefont {R.~R.}\ \bibnamefont {Rote}}, \bibinfo {author}
  {\bibfnamefont {B.}~\bibnamefont {T\'egl\'as}}, \bibinfo {author}
  {\bibfnamefont {C.}~\bibnamefont {Unsworth}}, \ and\ \bibinfo {author}
  {\bibfnamefont {C.}~\bibnamefont {Zhang}},\ }\href@noop {} {\bibfield
  {journal} {\bibinfo  {journal} {arXiv e-prints}\ } (\bibinfo {year}
  {2021}{\natexlab{b}})},\ \Eprint {http://arxiv.org/abs/2103.12104}
  {arXiv:2103.12104 [gr-qc]} \BibitemShut {NoStop}%
\bibitem [{\citenamefont {Cabero}\ \emph {et~al.}(2019)\citenamefont {Cabero},
  \citenamefont {Lundgren}, \citenamefont {Nitz}, \citenamefont {Dent},
  \citenamefont {Barker}, \citenamefont {Goetz}, \citenamefont {Kissel},
  \citenamefont {Nuttall}, \citenamefont {Schale}, \citenamefont {Schofield},\
  and\ \citenamefont {et~al.}}]{Cabero_2019}%
  \BibitemOpen
  \bibfield  {author} {\bibinfo {author} {\bibfnamefont {M.}~\bibnamefont
  {Cabero}}, \bibinfo {author} {\bibfnamefont {A.}~\bibnamefont {Lundgren}},
  \bibinfo {author} {\bibfnamefont {A.~H.}\ \bibnamefont {Nitz}}, \bibinfo
  {author} {\bibfnamefont {T.}~\bibnamefont {Dent}}, \bibinfo {author}
  {\bibfnamefont {D.}~\bibnamefont {Barker}}, \bibinfo {author} {\bibfnamefont
  {E.}~\bibnamefont {Goetz}}, \bibinfo {author} {\bibfnamefont {J.~S.}\
  \bibnamefont {Kissel}}, \bibinfo {author} {\bibfnamefont {L.~K.}\
  \bibnamefont {Nuttall}}, \bibinfo {author} {\bibfnamefont {P.}~\bibnamefont
  {Schale}}, \bibinfo {author} {\bibfnamefont {R.}~\bibnamefont {Schofield}}, \
  and\ \bibinfo {author} {\bibnamefont {et~al.}},\ }\href {\doibase
  10.1088/1361-6382/ab2e14} {\bibfield  {journal} {\bibinfo  {journal} {Class.
  Quant. Grav.}\ }\textbf {\bibinfo {volume} {36}},\ \bibinfo {pages} {155010}
  (\bibinfo {year} {2019})}\BibitemShut {NoStop}%
\bibitem [{\citenamefont {Klimenko}\ \emph {et~al.}(2015)\citenamefont
  {Klimenko}, \citenamefont {Vedovato}, \citenamefont {Drago}, \citenamefont
  {Salemi}, \citenamefont {Tiwari}, \citenamefont {Prodi}, \citenamefont
  {Lazzaro}, \citenamefont {Tiwari}, \citenamefont {Da~Silva},\ and\
  \citenamefont {Mitselmakher}}]{cwb2g}%
  \BibitemOpen
  \bibfield  {author} {\bibinfo {author} {\bibfnamefont {S.}~\bibnamefont
  {Klimenko}}, \bibinfo {author} {\bibfnamefont {G.}~\bibnamefont {Vedovato}},
  \bibinfo {author} {\bibfnamefont {M.}~\bibnamefont {Drago}}, \bibinfo
  {author} {\bibfnamefont {F.}~\bibnamefont {Salemi}}, \bibinfo {author}
  {\bibfnamefont {V.}~\bibnamefont {Tiwari}}, \bibinfo {author} {\bibfnamefont
  {G.~A.}\ \bibnamefont {Prodi}}, \bibinfo {author} {\bibfnamefont
  {C.}~\bibnamefont {Lazzaro}}, \bibinfo {author} {\bibfnamefont
  {S.}~\bibnamefont {Tiwari}}, \bibinfo {author} {\bibfnamefont
  {F.}~\bibnamefont {Da~Silva}}, \ and\ \bibinfo {author} {\bibfnamefont
  {G.}~\bibnamefont {Mitselmakher}},\ }\href
  {https://journals.aps.org/prd/abstract/10.1103/PhysRevD.93.042004} {\bibfield
   {journal} {\bibinfo  {journal} {Phys. Rev. D 93, 042004}\ } (\bibinfo {year}
  {2015})}\BibitemShut {NoStop}%
%%CITATION = ARXIV:1511.05999;%%
\bibitem [{\citenamefont {Drago}\ \emph {et~al.}(2021)\citenamefont {Drago}
  \emph {et~al.}}]{SOFTX}%
  \BibitemOpen
  \bibfield  {author} {\bibinfo {author} {\bibfnamefont {M.}~\bibnamefont
  {Drago}} \emph {et~al.},\ }\href {\doibase
  https://doi.org/10.1016/j.softx.2021.100678} {\bibfield  {journal} {\bibinfo
  {journal} {SoftwareX}\ }\textbf {\bibinfo {volume} {14}},\ \bibinfo {pages}
  {100678} (\bibinfo {year} {2021})}\BibitemShut {NoStop}%
\bibitem [{cwb()}]{cwb_repo}%
  \BibitemOpen
  \href@noop {} {}\bibinfo {howpublished}
  {\url{https://gwburst.gitlab.io}}\BibitemShut {NoStop}%
\bibitem [{\citenamefont {Necula}\ \emph {et~al.}(2012)\citenamefont {Necula},
  \citenamefont {Klimenko},\ and\ \citenamefont {Mitselmakher}}]{WDM}%
  \BibitemOpen
  \bibfield  {author} {\bibinfo {author} {\bibfnamefont {V.}~\bibnamefont
  {Necula}}, \bibinfo {author} {\bibfnamefont {S.}~\bibnamefont {Klimenko}}, \
  and\ \bibinfo {author} {\bibfnamefont {G.}~\bibnamefont {Mitselmakher}},\
  }\href {\doibase 10.1088/1742-6596/363/1/012032} {\bibfield  {journal}
  {\bibinfo  {journal} {J. Phys. Conf. Ser.}\ }\textbf {\bibinfo {volume}
  {363}},\ \bibinfo {pages} {012032} (\bibinfo {year} {2012})}\BibitemShut
  {NoStop}%
\bibitem [{\citenamefont {Cornish}\ and\ \citenamefont
  {Littenberg}(2015)}]{Cornish:2014kda}%
  \BibitemOpen
  \bibfield  {author} {\bibinfo {author} {\bibfnamefont {N.}~\bibnamefont
  {Cornish}}\ and\ \bibinfo {author} {\bibfnamefont {T.}~\bibnamefont
  {Littenberg}},\ }\href {\doibase 10.1088/0264-9381/32/13/135012} {\bibfield
  {journal} {\bibinfo  {journal} {Class. Quant. Grav.}\ }\textbf {\bibinfo
  {volume} {32}},\ \bibinfo {pages} {135012} (\bibinfo {year}
  {2015})}\BibitemShut {NoStop}%
%%CITATION = ARXIV:1410.3835;%%
\bibitem [{bw_()}]{bw_repo}%
  \BibitemOpen
  \href@noop {} {}\bibinfo {howpublished}
  {\url{https://git.ligo.org/lscsoft/bayeswave}}\BibitemShut {NoStop}%
\bibitem [{\citenamefont {{Cornish}}\ \emph {et~al.}(2021)\citenamefont
  {{Cornish}}, \citenamefont {{Littenberg}}, \citenamefont {{B{\'e}csy}},
  \citenamefont {{Chatziioannou}}, \citenamefont {{Clark}}, \citenamefont
  {{Ghonge}},\ and\ \citenamefont {{Millhouse}}}]{BayesWaveMethodsIII}%
  \BibitemOpen
  \bibfield  {author} {\bibinfo {author} {\bibfnamefont {N.~J.}\ \bibnamefont
  {{Cornish}}}, \bibinfo {author} {\bibfnamefont {T.~B.}\ \bibnamefont
  {{Littenberg}}}, \bibinfo {author} {\bibfnamefont {B.}~\bibnamefont
  {{B{\'e}csy}}}, \bibinfo {author} {\bibfnamefont {K.}~\bibnamefont
  {{Chatziioannou}}}, \bibinfo {author} {\bibfnamefont {J.~A.}\ \bibnamefont
  {{Clark}}}, \bibinfo {author} {\bibfnamefont {S.}~\bibnamefont {{Ghonge}}}, \
  and\ \bibinfo {author} {\bibfnamefont {M.}~\bibnamefont {{Millhouse}}},\
  }\href {http://dx.doi.org/10.1103/PhysRevD.103.044006} {\bibfield  {journal}
  {\bibinfo  {journal} {Phys. Rev. D}\ }\textbf {\bibinfo {volume} {103}},\
  \bibinfo {pages} {044006} (\bibinfo {year} {2021})}\BibitemShut {NoStop}%
\bibitem [{\citenamefont {{LIGO Scientific Collaboration}}\ and\ \citenamefont
  {{Virgo Collaboration}}(2019{\natexlab{a}})}]{GCNS190928c}%
  \BibitemOpen
  \bibfield  {author} {\bibinfo {author} {\bibnamefont {{LIGO Scientific
  Collaboration}}}\ and\ \bibinfo {author} {\bibnamefont {{Virgo
  Collaboration}}},\ }\href {https://gcn.gsfc.nasa.gov/other/GW190928c.gcn3}
  {\bibfield  {journal} {\bibinfo  {journal} {GCN}\ }\textbf {\bibinfo {volume}
  {25883}} (\bibinfo {year} {2019}{\natexlab{a}})}\BibitemShut {NoStop}%
\bibitem [{\citenamefont {Zevin}\ \emph {et~al.}(2017)\citenamefont {Zevin},
  \citenamefont {Coughlin}, \citenamefont {Bahaadini}, \citenamefont {Besler},
  \citenamefont {Rohani}, \citenamefont {Allen}, \citenamefont {Cabero},
  \citenamefont {Crowston}, \citenamefont {Katsaggelos}, \citenamefont
  {Larson}, \citenamefont {Lee}, \citenamefont {Lintott}, \citenamefont
  {Littenberg}, \citenamefont {Lundgren}, \citenamefont {Osterlund},
  \citenamefont {Smith}, \citenamefont {Trouille},\ and\ \citenamefont
  {Kalogera}}]{gravityspy}%
  \BibitemOpen
  \bibfield  {author} {\bibinfo {author} {\bibfnamefont {M.}~\bibnamefont
  {Zevin}}, \bibinfo {author} {\bibfnamefont {S.}~\bibnamefont {Coughlin}},
  \bibinfo {author} {\bibfnamefont {S.}~\bibnamefont {Bahaadini}}, \bibinfo
  {author} {\bibfnamefont {E.}~\bibnamefont {Besler}}, \bibinfo {author}
  {\bibfnamefont {N.}~\bibnamefont {Rohani}}, \bibinfo {author} {\bibfnamefont
  {S.}~\bibnamefont {Allen}}, \bibinfo {author} {\bibfnamefont
  {M.}~\bibnamefont {Cabero}}, \bibinfo {author} {\bibfnamefont
  {K.}~\bibnamefont {Crowston}}, \bibinfo {author} {\bibfnamefont
  {A.}~\bibnamefont {Katsaggelos}}, \bibinfo {author} {\bibfnamefont
  {S.}~\bibnamefont {Larson}}, \bibinfo {author} {\bibfnamefont
  {T.}~\bibnamefont {Lee}}, \bibinfo {author} {\bibfnamefont {C.}~\bibnamefont
  {Lintott}}, \bibinfo {author} {\bibfnamefont {T.}~\bibnamefont {Littenberg}},
  \bibinfo {author} {\bibfnamefont {A.}~\bibnamefont {Lundgren}}, \bibinfo
  {author} {\bibfnamefont {C.}~\bibnamefont {Osterlund}}, \bibinfo {author}
  {\bibfnamefont {J.}~\bibnamefont {Smith}}, \bibinfo {author} {\bibfnamefont
  {L.}~\bibnamefont {Trouille}}, \ and\ \bibinfo {author} {\bibfnamefont
  {V.}~\bibnamefont {Kalogera}},\ }\href
  {https://iopscience.iop.org/article/10.1088/1361-6382/aa5cea} {\bibfield
  {journal} {\bibinfo  {journal} {Class. Quant. Grav.}\ }\textbf {\bibinfo
  {volume} {34}},\ \bibinfo {pages} {103009} (\bibinfo {year}
  {2017})}\BibitemShut {NoStop}%
\bibitem [{\citenamefont {{LIGO Scientific Collaboration}}\ and\ \citenamefont
  {{Virgo Collaboration}}(2019{\natexlab{b}})}]{GCNS191110af}%
  \BibitemOpen
  \bibfield  {author} {\bibinfo {author} {\bibnamefont {{LIGO Scientific
  Collaboration}}}\ and\ \bibinfo {author} {\bibnamefont {{Virgo
  Collaboration}}},\ }\href {https://gcn.gsfc.nasa.gov/other/GW191110af.gcn3}
  {\bibfield  {journal} {\bibinfo  {journal} {GCN}\ }\textbf {\bibinfo {volume}
  {26220}} (\bibinfo {year} {2019}{\natexlab{b}})}\BibitemShut {NoStop}%
\bibitem [{\citenamefont {{LIGO Scientific Collaboration}}\ and\ \citenamefont
  {{Virgo Collaboration}}(2020)}]{GCNS200114f}%
  \BibitemOpen
  \bibfield  {author} {\bibinfo {author} {\bibnamefont {{LIGO Scientific
  Collaboration}}}\ and\ \bibinfo {author} {\bibnamefont {{Virgo
  Collaboration}}},\ }\href {https://gcn.gsfc.nasa.gov/other/GW200114f.gcn3}
  {\bibfield  {journal} {\bibinfo  {journal} {GCN}\ }\textbf {\bibinfo {volume}
  {26731}} (\bibinfo {year} {2020})}\BibitemShut {NoStop}%
\bibitem [{\citenamefont {Abadie}\ \emph {et~al.}(2012)\citenamefont {Abadie}
  \emph {et~al.}}]{s6Burst}%
  \BibitemOpen
  \bibfield  {author} {\bibinfo {author} {\bibfnamefont {J.}~\bibnamefont
  {Abadie}} \emph {et~al.} (\bibinfo {collaboration} {LIGO Scientific
  Collaboration and Virgo Collaboration}),\ }\href {\doibase
  10.1103/PhysRevD.85.122007} {\bibfield  {journal} {\bibinfo  {journal}
  {\prd}\ }\textbf {\bibinfo {volume} {85}},\ \bibinfo {eid} {122007} (\bibinfo
  {year} {2012})}\BibitemShut {NoStop}%
\bibitem [{\citenamefont {Kanner}\ \emph {et~al.}(2016)\citenamefont {Kanner},
  \citenamefont {Littenberg}, \citenamefont {Cornish}, \citenamefont
  {Millhouse}, \citenamefont {Xhakaj}, \citenamefont {Salemi}, \citenamefont
  {Drago}, \citenamefont {Vedovato},\ and\ \citenamefont
  {Klimenko}}]{Kanner:2015xua}%
  \BibitemOpen
  \bibfield  {author} {\bibinfo {author} {\bibfnamefont {J.~B.}\ \bibnamefont
  {Kanner}}, \bibinfo {author} {\bibfnamefont {T.~B.}\ \bibnamefont
  {Littenberg}}, \bibinfo {author} {\bibfnamefont {N.}~\bibnamefont {Cornish}},
  \bibinfo {author} {\bibfnamefont {M.}~\bibnamefont {Millhouse}}, \bibinfo
  {author} {\bibfnamefont {E.}~\bibnamefont {Xhakaj}}, \bibinfo {author}
  {\bibfnamefont {F.}~\bibnamefont {Salemi}}, \bibinfo {author} {\bibfnamefont
  {M.}~\bibnamefont {Drago}}, \bibinfo {author} {\bibfnamefont
  {G.}~\bibnamefont {Vedovato}}, \ and\ \bibinfo {author} {\bibfnamefont
  {S.}~\bibnamefont {Klimenko}},\ }\href {\doibase 10.1103/PhysRevD.93.022002}
  {\bibfield  {journal} {\bibinfo  {journal} {Phys. Rev. D}\ }\textbf {\bibinfo
  {volume} {93}},\ \bibinfo {pages} {022002} (\bibinfo {year}
  {2016})}\BibitemShut {NoStop}%
%%CITATION = ARXIV:1509.06423;%%
\bibitem [{\citenamefont {{Lee}}\ \emph {et~al.}(2021)\citenamefont {{Lee}},
  \citenamefont {{Millhouse}},\ and\ \citenamefont {{Melatos}}}]{Lee:2021}%
  \BibitemOpen
  \bibfield  {author} {\bibinfo {author} {\bibfnamefont {Y.~S.~C.}\
  \bibnamefont {{Lee}}}, \bibinfo {author} {\bibfnamefont {M.}~\bibnamefont
  {{Millhouse}}}, \ and\ \bibinfo {author} {\bibfnamefont {A.}~\bibnamefont
  {{Melatos}}},\ }\href {\doibase 10.1103/PhysRevD.103.062002} {\bibfield
  {journal} {\bibinfo  {journal} {\prd}\ }\textbf {\bibinfo {volume} {103}},\
  \bibinfo {eid} {062002} (\bibinfo {year} {2021})}\BibitemShut {NoStop}%
\bibitem [{\citenamefont {Sutton}(2013)}]{sutton-rule}%
  \BibitemOpen
  \bibfield  {author} {\bibinfo {author} {\bibfnamefont {P.~J.}\ \bibnamefont
  {Sutton}},\ }\href@noop {} {\bibfield  {journal} {\bibinfo  {journal} {arXiv
  e-prints}\ } (\bibinfo {year} {2013})},\ \Eprint
  {http://arxiv.org/abs/1304.0210} {arXiv:1304.0210 [gr-qc]} \BibitemShut
  {NoStop}%
\bibitem [{\citenamefont {{Abdikamalov}}\ \emph {et~al.}(2020)\citenamefont
  {{Abdikamalov}}, \citenamefont {{Pagliaroli}},\ and\ \citenamefont
  {{Radice}}}]{SNGW_review}%
  \BibitemOpen
  \bibfield  {author} {\bibinfo {author} {\bibfnamefont {E.}~\bibnamefont
  {{Abdikamalov}}}, \bibinfo {author} {\bibfnamefont {G.}~\bibnamefont
  {{Pagliaroli}}}, \ and\ \bibinfo {author} {\bibfnamefont {D.}~\bibnamefont
  {{Radice}}},\ }\href@noop {} {\bibfield  {journal} {\bibinfo  {journal}
  {arXiv e-prints}\ } (\bibinfo {year} {2020})},\ \Eprint
  {http://arxiv.org/abs/2010.04356} {arXiv:2010.04356 [astro-ph.SR]}
  \BibitemShut {NoStop}%
\bibitem [{\citenamefont {{Abbott}}\ \emph {et~al.}(2016)\citenamefont
  {{Abbott}} \emph {et~al.}}]{targeted_SN_initialLIGO}%
  \BibitemOpen
  \bibfield  {author} {\bibinfo {author} {\bibfnamefont {B.~P.}\ \bibnamefont
  {{Abbott}}} \emph {et~al.},\ }\href {\doibase 10.1103/PhysRevD.94.102001}
  {\bibfield  {journal} {\bibinfo  {journal} {\prd}\ }\textbf {\bibinfo
  {volume} {94}},\ \bibinfo {eid} {102001} (\bibinfo {year}
  {2016})}\BibitemShut {NoStop}%
\bibitem [{\citenamefont {{O'Connor}}(2017)}]{failed_SN}%
  \BibitemOpen
  \bibfield  {author} {\bibinfo {author} {\bibfnamefont {E.}~\bibnamefont
  {{O'Connor}}},\ }\enquote {\bibinfo {title} {{The Core-Collapse
  Supernova-Black Hole Connection}},}\ in\ \href {\doibase
  10.1007/978-3-319-21846-5_129} {\emph {\bibinfo {booktitle} {Handbook of
  Supernovae}}},\ \bibinfo {editor} {edited by\ \bibinfo {editor}
  {\bibfnamefont {A.~W.}\ \bibnamefont {{Alsabti}}}\ and\ \bibinfo {editor}
  {\bibfnamefont {P.}~\bibnamefont {{Murdin}}}}\ (\bibinfo {year} {2017})\ p.\
  \bibinfo {pages} {1555}\BibitemShut {NoStop}%
\bibitem [{\citenamefont {Szczepa\'nczyk}\ \emph {et~al.}(2021)\citenamefont
  {Szczepa\'nczyk} \emph {et~al.}}]{Szczepanczyk:2021bka}%
  \BibitemOpen
  \bibfield  {author} {\bibinfo {author} {\bibfnamefont {M.~J.}\ \bibnamefont
  {Szczepa\'nczyk}} \emph {et~al.},\ }\href@noop {} {\bibfield  {journal}
  {\bibinfo  {journal} {arXiv e-prints}\ } (\bibinfo {year} {2021})},\ \Eprint
  {http://arxiv.org/abs/2104.06462} {arXiv:2104.06462 [astro-ph.HE]}
  \BibitemShut {NoStop}%
\bibitem [{\citenamefont {{Powell}}\ and\ \citenamefont
  {{M{\"u}ller}}(2019)}]{Powell_2018}%
  \BibitemOpen
  \bibfield  {author} {\bibinfo {author} {\bibfnamefont {J.}~\bibnamefont
  {{Powell}}}\ and\ \bibinfo {author} {\bibfnamefont {B.}~\bibnamefont
  {{M{\"u}ller}}},\ }\href {\doibase 10.1093/mnras/stz1304} {\bibfield
  {journal} {\bibinfo  {journal} {\mnras}\ }\textbf {\bibinfo {volume} {487}},\
  \bibinfo {pages} {1178} (\bibinfo {year} {2019})}\BibitemShut {NoStop}%
\bibitem [{\citenamefont {{O'Connor}}\ and\ \citenamefont
  {{Couch}}(2018)}]{OConnor_Couch_2018}%
  \BibitemOpen
  \bibfield  {author} {\bibinfo {author} {\bibfnamefont {E.~P.}\ \bibnamefont
  {{O'Connor}}}\ and\ \bibinfo {author} {\bibfnamefont {S.~M.}\ \bibnamefont
  {{Couch}}},\ }\href {\doibase 10.3847/1538-4357/aadcf7} {\bibfield  {journal}
  {\bibinfo  {journal} {Astrophys. J}\ }\textbf {\bibinfo {volume} {865}},\
  \bibinfo {eid} {81} (\bibinfo {year} {2018})}\BibitemShut {NoStop}%
\bibitem [{\citenamefont {Blondin}\ \emph {et~al.}(2003)\citenamefont
  {Blondin}, \citenamefont {Mezzacappa},\ and\ \citenamefont
  {DeMarino}}]{SASI1}%
  \BibitemOpen
  \bibfield  {author} {\bibinfo {author} {\bibfnamefont {J.~M.}\ \bibnamefont
  {Blondin}}, \bibinfo {author} {\bibfnamefont {A.}~\bibnamefont {Mezzacappa}},
  \ and\ \bibinfo {author} {\bibfnamefont {C.}~\bibnamefont {DeMarino}},\
  }\href {http://stacks.iop.org/0004-637X/584/i=2/a=971} {\bibfield  {journal}
  {\bibinfo  {journal} {Astrophys. J}\ }\textbf {\bibinfo {volume} {584}},\
  \bibinfo {pages} {971} (\bibinfo {year} {2003})}\BibitemShut {NoStop}%
\bibitem [{\citenamefont {{Blondin}}\ and\ \citenamefont
  {{Mezzacappa}}(2006)}]{SASI2}%
  \BibitemOpen
  \bibfield  {author} {\bibinfo {author} {\bibfnamefont {J.~M.}\ \bibnamefont
  {{Blondin}}}\ and\ \bibinfo {author} {\bibfnamefont {A.}~\bibnamefont
  {{Mezzacappa}}},\ }\href {\doibase 10.1086/500817} {\bibfield  {journal}
  {\bibinfo  {journal} {Astrophys. J}\ }\textbf {\bibinfo {volume} {642}},\
  \bibinfo {pages} {401} (\bibinfo {year} {2006})}\BibitemShut {NoStop}%
\bibitem [{\citenamefont {{Foglizzo}}\ \emph {et~al.}(2007)\citenamefont
  {{Foglizzo}}, \citenamefont {{Galletti}}, \citenamefont {{Scheck}},\ and\
  \citenamefont {{Janka}}}]{SASI3}%
  \BibitemOpen
  \bibfield  {author} {\bibinfo {author} {\bibfnamefont {T.}~\bibnamefont
  {{Foglizzo}}}, \bibinfo {author} {\bibfnamefont {P.}~\bibnamefont
  {{Galletti}}}, \bibinfo {author} {\bibfnamefont {L.}~\bibnamefont
  {{Scheck}}}, \ and\ \bibinfo {author} {\bibfnamefont {H.-T.}\ \bibnamefont
  {{Janka}}},\ }\href {\doibase 10.1086/509612} {\bibfield  {journal} {\bibinfo
   {journal} {Astrophys. J}\ }\textbf {\bibinfo {volume} {654}},\ \bibinfo
  {pages} {1006} (\bibinfo {year} {2007})}\BibitemShut {NoStop}%
\bibitem [{\citenamefont {{Radice}}\ \emph {et~al.}(2019)\citenamefont
  {{Radice}}, \citenamefont {{Morozova}}, \citenamefont {{Burrows}},
  \citenamefont {{Vartanyan}},\ and\ \citenamefont {{Nagakura}}}]{Radice_2019}%
  \BibitemOpen
  \bibfield  {author} {\bibinfo {author} {\bibfnamefont {D.}~\bibnamefont
  {{Radice}}}, \bibinfo {author} {\bibfnamefont {V.}~\bibnamefont
  {{Morozova}}}, \bibinfo {author} {\bibfnamefont {A.}~\bibnamefont
  {{Burrows}}}, \bibinfo {author} {\bibfnamefont {D.}~\bibnamefont
  {{Vartanyan}}}, \ and\ \bibinfo {author} {\bibfnamefont {H.}~\bibnamefont
  {{Nagakura}}},\ }\href {\doibase 10.3847/2041-8213/ab191a} {\bibfield
  {journal} {\bibinfo  {journal} {Astrophys. J. Lett.}\ }\textbf {\bibinfo
  {volume} {876}},\ \bibinfo {eid} {L9} (\bibinfo {year} {2019})}\BibitemShut
  {NoStop}%
\bibitem [{\citenamefont {Janka}(2012)}]{Janka:2012wk}%
  \BibitemOpen
  \bibfield  {author} {\bibinfo {author} {\bibfnamefont {H.-T.}\ \bibnamefont
  {Janka}},\ }\href {\doibase 10.1146/annurev-nucl-102711-094901} {\bibfield
  {journal} {\bibinfo  {journal} {Ann. Rev. Nucl. Part. Sci.}\ }\textbf
  {\bibinfo {volume} {62}},\ \bibinfo {pages} {407} (\bibinfo {year} {2012})},\
  \Eprint {http://arxiv.org/abs/1206.2503} {1206.2503} \BibitemShut {NoStop}%
\bibitem [{\citenamefont {Woosley}\ and\ \citenamefont
  {Heger}(2006)}]{Woosley:2005gy}%
  \BibitemOpen
  \bibfield  {author} {\bibinfo {author} {\bibfnamefont {S.}~\bibnamefont
  {Woosley}}\ and\ \bibinfo {author} {\bibfnamefont {A.}~\bibnamefont
  {Heger}},\ }\href {\doibase 10.1086/498500} {\bibfield  {journal} {\bibinfo
  {journal} {Astrophys. J.}\ }\textbf {\bibinfo {volume} {637}},\ \bibinfo
  {pages} {914} (\bibinfo {year} {2006})}\BibitemShut {NoStop}%
\bibitem [{\citenamefont {{Powell}}\ and\ \citenamefont
  {{M{\"u}ller}}(2020)}]{Powell_2020}%
  \BibitemOpen
  \bibfield  {author} {\bibinfo {author} {\bibfnamefont {J.}~\bibnamefont
  {{Powell}}}\ and\ \bibinfo {author} {\bibfnamefont {B.}~\bibnamefont
  {{M{\"u}ller}}},\ }\href {\doibase 10.1093/mnras/staa1048} {\bibfield
  {journal} {\bibinfo  {journal} {\mnras}\ }\textbf {\bibinfo {volume} {494}},\
  \bibinfo {pages} {4665} (\bibinfo {year} {2020})}\BibitemShut {NoStop}%
\bibitem [{\citenamefont {{Obergaulinger}}\ and\ \citenamefont
  {{Aloy}}(2020)}]{Obergaulinger_2020}%
  \BibitemOpen
  \bibfield  {author} {\bibinfo {author} {\bibfnamefont {M.}~\bibnamefont
  {{Obergaulinger}}}\ and\ \bibinfo {author} {\bibfnamefont {M.~{\'A}.}\
  \bibnamefont {{Aloy}}},\ }\href {\doibase 10.1093/mnras/staa096} {\bibfield
  {journal} {\bibinfo  {journal} {\mnras}\ }\textbf {\bibinfo {volume} {492}},\
  \bibinfo {pages} {4613} (\bibinfo {year} {2020})}\BibitemShut {NoStop}%
\bibitem [{\citenamefont {{Abe}}\ \emph {et~al.}(2016)\citenamefont {{Abe}},
  \citenamefont {{Haga}}, \citenamefont {{Hayato}}, \citenamefont {{Ikeda}},
  \citenamefont {{Iyogi}}, \citenamefont {{Kameda}}, \citenamefont
  {{Kishimoto}}, \citenamefont {{Miura}}, \citenamefont {{Moriyama}},
  \citenamefont {{Nakahata}}, \citenamefont {{Nakano}}, \citenamefont
  {{Nakayama}}, \citenamefont {{Sekiya}}, \citenamefont {{Shiozawa}},
  \citenamefont {{Suzuki}} \emph {et~al.}}]{SK_sensitivity}%
  \BibitemOpen
  \bibfield  {author} {\bibinfo {author} {\bibfnamefont {K.}~\bibnamefont
  {{Abe}}}, \bibinfo {author} {\bibfnamefont {Y.}~\bibnamefont {{Haga}}},
  \bibinfo {author} {\bibfnamefont {Y.}~\bibnamefont {{Hayato}}}, \bibinfo
  {author} {\bibfnamefont {M.}~\bibnamefont {{Ikeda}}}, \bibinfo {author}
  {\bibfnamefont {K.}~\bibnamefont {{Iyogi}}}, \bibinfo {author} {\bibfnamefont
  {J.}~\bibnamefont {{Kameda}}}, \bibinfo {author} {\bibfnamefont
  {Y.}~\bibnamefont {{Kishimoto}}}, \bibinfo {author} {\bibfnamefont
  {M.}~\bibnamefont {{Miura}}}, \bibinfo {author} {\bibfnamefont
  {S.}~\bibnamefont {{Moriyama}}}, \bibinfo {author} {\bibfnamefont
  {M.}~\bibnamefont {{Nakahata}}}, \bibinfo {author} {\bibfnamefont
  {Y.}~\bibnamefont {{Nakano}}}, \bibinfo {author} {\bibfnamefont
  {S.}~\bibnamefont {{Nakayama}}}, \bibinfo {author} {\bibfnamefont
  {H.}~\bibnamefont {{Sekiya}}}, \bibinfo {author} {\bibfnamefont
  {M.}~\bibnamefont {{Shiozawa}}}, \bibinfo {author} {\bibnamefont {{Suzuki}}},
   \emph {et~al.},\ }\href {\doibase 10.1016/j.astropartphys.2016.04.003}
  {\bibfield  {journal} {\bibinfo  {journal} {Astroparticle Physics}\ }\textbf
  {\bibinfo {volume} {81}},\ \bibinfo {pages} {39} (\bibinfo {year}
  {2016})}\BibitemShut {NoStop}%
\bibitem [{\citenamefont {{McMillan}}(2017)}]{MilkyWay1}%
  \BibitemOpen
  \bibfield  {author} {\bibinfo {author} {\bibfnamefont {P.~J.}\ \bibnamefont
  {{McMillan}}},\ }\href {\doibase 10.1093/mnras/stw2759} {\bibfield  {journal}
  {\bibinfo  {journal} {\mnras}\ }\textbf {\bibinfo {volume} {465}},\ \bibinfo
  {pages} {76} (\bibinfo {year} {2017})}\BibitemShut {NoStop}%
\bibitem [{\citenamefont {{Cautun}}\ \emph {et~al.}(2020)\citenamefont
  {{Cautun}}, \citenamefont {{Ben{\'\i}tez-Llambay}}, \citenamefont {{Deason}},
  \citenamefont {{Frenk}}, \citenamefont {{Fattahi}}, \citenamefont
  {{G{\'o}mez}}, \citenamefont {{Grand}}, \citenamefont {{Oman}}, \citenamefont
  {{Navarro}},\ and\ \citenamefont {{Simpson}}}]{MilkyWay2}%
  \BibitemOpen
  \bibfield  {author} {\bibinfo {author} {\bibfnamefont {M.}~\bibnamefont
  {{Cautun}}}, \bibinfo {author} {\bibfnamefont {A.}~\bibnamefont
  {{Ben{\'\i}tez-Llambay}}}, \bibinfo {author} {\bibfnamefont {A.~J.}\
  \bibnamefont {{Deason}}}, \bibinfo {author} {\bibfnamefont {C.~S.}\
  \bibnamefont {{Frenk}}}, \bibinfo {author} {\bibfnamefont {A.}~\bibnamefont
  {{Fattahi}}}, \bibinfo {author} {\bibfnamefont {F.~A.}\ \bibnamefont
  {{G{\'o}mez}}}, \bibinfo {author} {\bibfnamefont {R.~J.~J.}\ \bibnamefont
  {{Grand}}}, \bibinfo {author} {\bibfnamefont {K.~A.}\ \bibnamefont {{Oman}}},
  \bibinfo {author} {\bibfnamefont {J.~F.}\ \bibnamefont {{Navarro}}}, \ and\
  \bibinfo {author} {\bibfnamefont {C.~M.}\ \bibnamefont {{Simpson}}},\ }\href
  {\doibase 10.1093/mnras/staa1017} {\bibfield  {journal} {\bibinfo  {journal}
  {\mnras}\ }\textbf {\bibinfo {volume} {494}},\ \bibinfo {pages} {4291}
  (\bibinfo {year} {2020})}\BibitemShut {NoStop}%
\bibitem [{\citenamefont {Abbott}\ \emph
  {et~al.}(2019{\natexlab{d}})\citenamefont {Abbott} \emph {et~al.}}]{Mag}%
  \BibitemOpen
  \bibfield  {author} {\bibinfo {author} {\bibfnamefont {B.~P.}\ \bibnamefont
  {Abbott}} \emph {et~al.} (\bibinfo {collaboration} {LIGO Scientific
  Collaboration and Virgo Collaboration}),\ }\href {\doibase
  10.3847/1538-4357/ab0e15} {\bibfield  {journal} {\bibinfo  {journal}
  {Astrophys. J.}\ }\textbf {\bibinfo {volume} {874}},\ \bibinfo {pages} {163}
  (\bibinfo {year} {2019}{\natexlab{d}})}\BibitemShut {NoStop}%
\bibitem [{\citenamefont {{Haskell}}\ and\ \citenamefont
  {{Melatos}}(2015)}]{mech_NS}%
  \BibitemOpen
  \bibfield  {author} {\bibinfo {author} {\bibfnamefont {B.}~\bibnamefont
  {{Haskell}}}\ and\ \bibinfo {author} {\bibfnamefont {A.}~\bibnamefont
  {{Melatos}}},\ }\href {\doibase 10.1142/S0218271815300086} {\bibfield
  {journal} {\bibinfo  {journal} {International Journal of Modern Physics D}\
  }\textbf {\bibinfo {volume} {24}},\ \bibinfo {eid} {1530008} (\bibinfo {year}
  {2015})}\BibitemShut {NoStop}%
\bibitem [{\citenamefont {Andersson}\ \emph {et~al.}(2004)\citenamefont
  {Andersson}, \citenamefont {Comer},\ and\ \citenamefont {Prix}}]{superfluid}%
  \BibitemOpen
  \bibfield  {author} {\bibinfo {author} {\bibfnamefont {N.}~\bibnamefont
  {Andersson}}, \bibinfo {author} {\bibfnamefont {G.~L.}\ \bibnamefont
  {Comer}}, \ and\ \bibinfo {author} {\bibfnamefont {R.}~\bibnamefont {Prix}},\
  }\href {\doibase 10.1111/j.1365-2966.2004.08166.x} {\bibfield  {journal}
  {\bibinfo  {journal} {\mnras}\ }\textbf {\bibinfo {volume} {354}},\ \bibinfo
  {pages} {101} (\bibinfo {year} {2004})}\BibitemShut {NoStop}%
\bibitem [{\citenamefont {Warszawski}\ and\ \citenamefont
  {Melatos}(2012)}]{sf2}%
  \BibitemOpen
  \bibfield  {author} {\bibinfo {author} {\bibfnamefont {L.}~\bibnamefont
  {Warszawski}}\ and\ \bibinfo {author} {\bibfnamefont {A.}~\bibnamefont
  {Melatos}},\ }\href {\doibase 10.1111/j.1365-2966.2012.20977.x} {\bibfield
  {journal} {\bibinfo  {journal} {\mnras}\ }\textbf {\bibinfo {volume} {423}},\
  \bibinfo {pages} {2058} (\bibinfo {year} {2012})}\BibitemShut {NoStop}%
\bibitem [{\citenamefont {Ho}\ \emph {et~al.}(2020)\citenamefont {Ho},
  \citenamefont {Jones}, \citenamefont {Andersson},\ and\ \citenamefont
  {Espinoza}}]{WH}%
  \BibitemOpen
  \bibfield  {author} {\bibinfo {author} {\bibfnamefont {W.~C.~G.}\
  \bibnamefont {Ho}}, \bibinfo {author} {\bibfnamefont {D.~I.}\ \bibnamefont
  {Jones}}, \bibinfo {author} {\bibfnamefont {N.}~\bibnamefont {Andersson}}, \
  and\ \bibinfo {author} {\bibfnamefont {C.~M.}\ \bibnamefont {Espinoza}},\
  }\href {\doibase 10.1103/PhysRevD.101.103009} {\bibfield  {journal} {\bibinfo
   {journal} {Phys. Rev. D}\ }\textbf {\bibinfo {volume} {101}},\ \bibinfo
  {pages} {103009} (\bibinfo {year} {2020})}\BibitemShut {NoStop}%
\bibitem [{\citenamefont {Caraveo}\ \emph {et~al.}(2001)\citenamefont
  {Caraveo}, \citenamefont {Luca},\ and\ \citenamefont {Mignani}}]{vdist}%
  \BibitemOpen
  \bibfield  {author} {\bibinfo {author} {\bibfnamefont {P.}~\bibnamefont
  {Caraveo}}, \bibinfo {author} {\bibfnamefont {A.}~\bibnamefont {Luca}}, \
  and\ \bibinfo {author} {\bibfnamefont {R.}~\bibnamefont {Mignani}},\ }\href
  {https://iopscience.iop.org/article/10.1086/323377} {\bibfield  {journal}
  {\bibinfo  {journal} {Astrophys. J.}\ }\textbf {\bibinfo {volume} {561}},\
  \bibinfo {pages} {930} (\bibinfo {year} {2001})}\BibitemShut {NoStop}%
\bibitem [{\citenamefont {{Manchester}}\ \emph {et~al.}(2005)\citenamefont
  {{Manchester}}, \citenamefont {{Hobbs}}, \citenamefont {{Teoh}},\ and\
  \citenamefont {{Hobbs}}}]{vspin}%
  \BibitemOpen
  \bibfield  {author} {\bibinfo {author} {\bibfnamefont {R.~N.}\ \bibnamefont
  {{Manchester}}}, \bibinfo {author} {\bibfnamefont {G.~B.}\ \bibnamefont
  {{Hobbs}}}, \bibinfo {author} {\bibfnamefont {A.}~\bibnamefont {{Teoh}}}, \
  and\ \bibinfo {author} {\bibfnamefont {M.}~\bibnamefont {{Hobbs}}},\ }\href
  {https://ui.adsabs.harvard.edu/abs/2005yCat.7245....0M/abstract} {\bibfield
  {journal} {\bibinfo  {journal} {{VizieR Online Data Catalog, VII/245}}\ }
  (\bibinfo {year} {2005})}\BibitemShut {NoStop}%
\bibitem [{\citenamefont {Doneva}\ \emph {et~al.}(2013)\citenamefont {Doneva},
  \citenamefont {Gaertig}, \citenamefont {Kokkotas},\ and\ \citenamefont
  {Kr\"uger}}]{MR-FT}%
  \BibitemOpen
  \bibfield  {author} {\bibinfo {author} {\bibfnamefont {D.~D.}\ \bibnamefont
  {Doneva}}, \bibinfo {author} {\bibfnamefont {E.}~\bibnamefont {Gaertig}},
  \bibinfo {author} {\bibfnamefont {K.~D.}\ \bibnamefont {Kokkotas}}, \ and\
  \bibinfo {author} {\bibfnamefont {C.}~\bibnamefont {Kr\"uger}},\ }\href
  {\doibase 10.1103/PhysRevD.88.044052} {\bibfield  {journal} {\bibinfo
  {journal} {Phys. Rev. D}\ }\textbf {\bibinfo {volume} {88}},\ \bibinfo
  {pages} {044052} (\bibinfo {year} {2013})}\BibitemShut {NoStop}%
\bibitem [{\citenamefont {Akmal}\ \emph {et~al.}(1998)\citenamefont {Akmal},
  \citenamefont {Pandharipande},\ and\ \citenamefont {Ravenhall}}]{apr4}%
  \BibitemOpen
  \bibfield  {author} {\bibinfo {author} {\bibfnamefont {A.}~\bibnamefont
  {Akmal}}, \bibinfo {author} {\bibfnamefont {V.~R.}\ \bibnamefont
  {Pandharipande}}, \ and\ \bibinfo {author} {\bibfnamefont {D.~G.}\
  \bibnamefont {Ravenhall}},\ }\href {\doibase 10.1103/PhysRevC.58.1804}
  {\bibfield  {journal} {\bibinfo  {journal} {Phys. Rev. C}\ }\textbf {\bibinfo
  {volume} {58}},\ \bibinfo {pages} {1804} (\bibinfo {year}
  {1998})}\BibitemShut {NoStop}%
\bibitem [{\citenamefont {Lackey}\ \emph {et~al.}(2006)\citenamefont {Lackey},
  \citenamefont {Nayyar},\ and\ \citenamefont {Owen}}]{h4}%
  \BibitemOpen
  \bibfield  {author} {\bibinfo {author} {\bibfnamefont {B.~D.}\ \bibnamefont
  {Lackey}}, \bibinfo {author} {\bibfnamefont {M.}~\bibnamefont {Nayyar}}, \
  and\ \bibinfo {author} {\bibfnamefont {B.~J.}\ \bibnamefont {Owen}},\ }\href
  {\doibase 10.1103/PhysRevD.73.024021} {\bibfield  {journal} {\bibinfo
  {journal} {Phys. Rev. D}\ }\textbf {\bibinfo {volume} {73}},\ \bibinfo
  {pages} {024021} (\bibinfo {year} {2006})}\BibitemShut {NoStop}%
\bibitem [{\citenamefont {Abbott}\ \emph
  {et~al.}(2018{\natexlab{b}})\citenamefont {Abbott} \emph
  {et~al.}}]{nsradius}%
  \BibitemOpen
  \bibfield  {author} {\bibinfo {author} {\bibfnamefont {B.~P.}\ \bibnamefont
  {Abbott}} \emph {et~al.} (\bibinfo {collaboration} {LIGO Scientific
  Collaboration and Virgo Collaboration}),\ }\href
  {https://journals.aps.org/prl/abstract/10.1103/PhysRevLett.121.161101}
  {\bibfield  {journal} {\bibinfo  {journal} {Phys. Rev. Lett.}\ }\textbf
  {\bibinfo {volume} {121}},\ \bibinfo {pages} {161101} (\bibinfo {year}
  {2018}{\natexlab{b}})}\BibitemShut {NoStop}%
\bibitem [{\citenamefont {Abbott}\ \emph
  {et~al.}(2020{\natexlab{a}})\citenamefont {Abbott} \emph
  {et~al.}}]{collab_eos_2}%
  \BibitemOpen
  \bibfield  {author} {\bibinfo {author} {\bibfnamefont {B.~P.}\ \bibnamefont
  {Abbott}} \emph {et~al.} (\bibinfo {collaboration} {LIGO Scientific
  Collaboration and Virgo Collaboration}),\ }\href {\doibase
  10.1088/1361-6382/ab5f7c} {\bibfield  {journal} {\bibinfo  {journal} {Class.
  Quant. Grav.}\ }\textbf {\bibinfo {volume} {37}},\ \bibinfo {pages} {045006}
  (\bibinfo {year} {2020}{\natexlab{a}})}\BibitemShut {NoStop}%
\bibitem [{\citenamefont {{Fuentes, J. R.}}\ \emph {et~al.}(2017)\citenamefont
  {{Fuentes, J. R.}}, \citenamefont {{Espinoza, C. M.}}, \citenamefont
  {{Reisenegger, A.}}, \citenamefont {{Shaw, B.}}, \citenamefont {{Stappers, B.
  W.}},\ and\ \citenamefont {{Lyne, A. G.}}}]{gref_1}%
  \BibitemOpen
  \bibfield  {author} {\bibinfo {author} {\bibnamefont {{Fuentes, J. R.}}},
  \bibinfo {author} {\bibnamefont {{Espinoza, C. M.}}}, \bibinfo {author}
  {\bibnamefont {{Reisenegger, A.}}}, \bibinfo {author} {\bibnamefont {{Shaw,
  B.}}}, \bibinfo {author} {\bibnamefont {{Stappers, B. W.}}}, \ and\ \bibinfo
  {author} {\bibnamefont {{Lyne, A. G.}}},\ }\href {\doibase
  10.1051/0004-6361/201731519} {\bibfield  {journal} {\bibinfo  {journal}
  {A\&A}\ }\textbf {\bibinfo {volume} {608}},\ \bibinfo {pages} {A131}
  (\bibinfo {year} {2017})}\BibitemShut {NoStop}%
\bibitem [{\citenamefont {Yu}\ \emph {et~al.}(2012)\citenamefont {Yu},
  \citenamefont {Manchester}, \citenamefont {Hobbs}, \citenamefont {Johnston},
  \citenamefont {Kaspi}, \citenamefont {Keith}, \citenamefont {Lyne},
  \citenamefont {Qiao}, \citenamefont {Ravi}, \citenamefont {Sarkissian},
  \citenamefont {Shannon},\ and\ \citenamefont {Xu}}]{gref_2}%
  \BibitemOpen
  \bibfield  {author} {\bibinfo {author} {\bibfnamefont {M.}~\bibnamefont
  {Yu}}, \bibinfo {author} {\bibfnamefont {R.~N.}\ \bibnamefont {Manchester}},
  \bibinfo {author} {\bibfnamefont {G.}~\bibnamefont {Hobbs}}, \bibinfo
  {author} {\bibfnamefont {S.}~\bibnamefont {Johnston}}, \bibinfo {author}
  {\bibfnamefont {V.~M.}\ \bibnamefont {Kaspi}}, \bibinfo {author}
  {\bibfnamefont {M.}~\bibnamefont {Keith}}, \bibinfo {author} {\bibfnamefont
  {A.~G.}\ \bibnamefont {Lyne}}, \bibinfo {author} {\bibfnamefont {G.~J.}\
  \bibnamefont {Qiao}}, \bibinfo {author} {\bibfnamefont {V.}~\bibnamefont
  {Ravi}}, \bibinfo {author} {\bibfnamefont {J.~M.}\ \bibnamefont
  {Sarkissian}}, \bibinfo {author} {\bibfnamefont {R.}~\bibnamefont {Shannon}},
  \ and\ \bibinfo {author} {\bibfnamefont {R.~X.}\ \bibnamefont {Xu}},\ }\href
  {\doibase 10.1093/mnras/sts366} {\bibfield  {journal} {\bibinfo  {journal}
  {\mnras}\ }\textbf {\bibinfo {volume} {429}},\ \bibinfo {pages} {688}
  (\bibinfo {year} {2012})}\BibitemShut {NoStop}%
\bibitem [{\citenamefont {Espinoza}\ \emph {et~al.}(2011)\citenamefont
  {Espinoza}, \citenamefont {Lyne}, \citenamefont {Stappers},\ and\
  \citenamefont {Kramer}}]{gref_3}%
  \BibitemOpen
  \bibfield  {author} {\bibinfo {author} {\bibfnamefont {C.~M.}\ \bibnamefont
  {Espinoza}}, \bibinfo {author} {\bibfnamefont {A.~G.}\ \bibnamefont {Lyne}},
  \bibinfo {author} {\bibfnamefont {B.~W.}\ \bibnamefont {Stappers}}, \ and\
  \bibinfo {author} {\bibfnamefont {M.}~\bibnamefont {Kramer}},\ }\href
  {\doibase 10.1111/j.1365-2966.2011.18503.x} {\bibfield  {journal} {\bibinfo
  {journal} {\mnras}\ }\textbf {\bibinfo {volume} {414}},\ \bibinfo {pages}
  {1679} (\bibinfo {year} {2011})}\BibitemShut {NoStop}%
\bibitem [{\citenamefont {Abbott}\ \emph
  {et~al.}(2020{\natexlab{b}})\citenamefont {Abbott} \emph
  {et~al.}}]{ObservingScenario}%
  \BibitemOpen
  \bibfield  {author} {\bibinfo {author} {\bibfnamefont {B.~P.}\ \bibnamefont
  {Abbott}} \emph {et~al.} (\bibinfo {collaboration} {LIGO Scientific
  Collaboration and Virgo Collaboration}),\ }\href
  {http://dx.doi.org/10.1007/s41114-020-00026-9} {\bibfield  {journal}
  {\bibinfo  {journal} {Living Reviews in Relativity}\ }\textbf {\bibinfo
  {volume} {23}} (\bibinfo {year} {2020}{\natexlab{b}})}\BibitemShut {NoStop}%
\end{thebibliography}%

% Uncomment to use the format where full author list comes at end of paper:
%  \let\author\myauthor
%  \let\affiliation\myaffiliation
%  \let\maketitle\mymaketitle
%  \title{Authors}
%  \input{LSC-Virgo-Authors-Feb-2019-prd}
%  \newpage
%  \maketitle

\end{document}